Growth, Characterization, and Thermodynamics of III-Nitride Semiconductors

by

Arlinda Hill

A Dissertation Presented in Partial Fulfillment
of the Requirements for the Degree
Doctor of Philosophy

Approved April 2011 by the
Graduate Supervisory Committee:

Fernando A. Ponce, Chair
Ralph V. Chamberlin
Otto F. Sankey
David J. Smith
Kong-Thon Tsen

ARIZONA STATE UNIVERSITY

May 2011

ABSTRACT

III-nitride alloys are wide band gap semiconductors with a broad range of applications in optoelectronic devices such as light emitting diodes and laser diodes. Indium gallium nitride light emitting diodes have been successfully produced over the past decade. But the progress of green emission light emitting devices has been limited by the incorporation of indium in the alloy, mainly due to phase separation. This difficulty could be addressed by studying the growth and thermodynamics of these alloys. Knowledge of thermodynamic phase stabilities and of pressure – temperature – composition phase diagrams is important for an understanding of the boundary conditions of a variety of growth techniques. In this dissertation a study of the phase separation of indium gallium nitride is conducted using a regular solution model of the ternary alloy system. Graphs of Gibbs free energy of mixing were produced for a range of temperatures. Binodal and spinodal decomposition curves show the stable and unstable regions of the alloy in equilibrium. The growth of gallium nitride and indium gallium nitride was attempted by the reaction of molten gallium – indium alloy with ammonia at atmospheric pressure. Characterization by X-ray diffraction, photoluminescence, and secondary electron microscopy show that the samples produced by this method contain only gallium nitride in the hexagonal phase. The instability of indium nitride at the temperatures required for activation of ammonia accounts for these results. The photoluminescence spectra show a correlation between the intensity of a broad green emission, related to native defects, and indium composition used in the molten alloy. A different growth method was used to grow two columnar-structured gallium nitride films using ammonium chloride and gallium as reactants and nitrogen and ammonia as carrier gasses. Investigation by X-ray diffraction and spatially-resolved cathodoluminescence shows the film grown at higher temperature to be primarily hexagonal with small quantities of cubic crystallites, while the one grown at lower temperature to be pure hexagonal. This was also confirmed by low temperature photoluminescence measurements. The results presented here show that cubic and hexagonal crystallites can coexist, with the cubic phase having a much sharper and stronger luminescence. Controlled growth of the cubic phase GaN crystallites can be of use for high efficiency light detecting and emitting devices.

The ammonolysis of a precursor was used to grow InGaN powders with different indium composition. High purity hexagonal GaN and InN were obtained. XRD spectra showed complete phase separation for samples with $x < 30\%$, with ~ 9% indium incorporation in the 30% sample. The presence of InGaN in this sample was confirmed by PL measurements, where luminescence from both GaN and InGaN band edge are observed. The growth of higher indium compositions samples proved to be difficult, with only the presence of InN in the sample. Nonetheless, by controlling parameters like temperature and time may lead to successful growth of this III-nitride alloy by this method.

To my loving husband,
my beautiful girls,
and my wonderful parents.

# ACKNOWLEDGEMENTS

I would like to start by thanking my advisor, Prof. Fernando Ponce, whose extensive knowledge, as well as his love of science, make him a truly exceptional mentor. I owe him a tremendous debt of gratitude for his guidance and support throughout my doctoral studies. My experiences with him will certainly guide me as I continue along in my career.

I would also like to thank my committee members Prof. Ralph Chamberlin, Prof. Otto Sankey, Prof. David Smith, and Prof. Kong-Ton Tsen. They have provided me with thoughtful insights, helpful comments and suggestions in this dissertation.

Thanks are also due to my research team members with whom I have had the pleasure to work and engage in discussions, namely Dr. Alec Fischer, Jingyi Huang, Yu Huang, Reid Juday, Ti Li, Kewei Sun, Travis Tobey, Qiyuan Wei, and Yong Wei. Special thanks go to Dr. Rafael Garcia and Mr. David Wright for their help with the growth process. I gratefully acknowledge the technical expertise and suggestions provided by Dr. Christian Poweleit, Dr. Thomas Groy, and Dr. Zhenquan Liu.

I thank the Department of Physics for the financial support offered to me in the form of teaching and research assistantships. I also acknowledge the kindness and helpfulness of faculty and staff members in the department.

Finally, I would like to thank all my family for their encouragement and unwavering support throughout my graduate studies. Your unending love was a tremendous driving force in the realization of this work.

TABLE OF CONTENTS

LIST OF TABLES

LIST OF FIGURES

# CHAPTER 1

# INTRODUCTION

## 1.1. GROUP III-NITRIDES

The III-nitrides and their alloys are direct bandgap semiconductors. This property makes them efficient emitters of light. Their bandgap varies from 0.7 eV for indium nitride (InN), 3.4 eV for gallium nitride (GaN), to 6.2 eV for aluminum nitride (AlN). Ternary III-nitride alloys like indium gallium nitride (InGaN), aluminum gallium nitride (AlGaN), or indium aluminum nitride (InAlN), offer the possibility for tuning of the bandgap of these alloys to correspond to light emission in the whole visible spectrum and into the deep ultraviolet (UV) region. As such, III-nitride alloys are particularly suitable for applications in optoelectronic devices for solid state lighting, such as laser diodes (LDs) and light emitting diodes (LEDs).[1] The wide bandgap energy range also makes these materials good candidates for absorber layers in solar cells since the absorption edge of these materials can be varied to optimize cell efficiency. III-nitride electronic devices are also more environmentally friendly because they do not contain toxic elements such as arsenic that are used to fabricate other compound semiconductors. High electron mobility and saturation velocity, high breakdown field, and high thermal conductivity are advantages of III-nitrides especially in view of applications for high-power, high-speed electronics.

Due to the great number of possible applications, III-nitrides have been studied vigorously; and in recent years they have gained a significant position in the science and technology of compound semiconductors. Gallium nitride (GaN) and its alloy indium gallium nitride (InGaN) have become dominant materials for producing high brightness light emitting diodes (LEDs) and laser diodes (LDs) that emit light in the green/blue region of the visible spectrum.  InGaN based light emitting diodes are already in use in full color liquid crystal diode (LCD) displays and traffic lights. All these applications demonstrate the technological relevance of the III-nitride compounds and the reason for these materials to be the subject of an active research field.

**1.2. BANDGAP ENERGY OF III-NITRIDES**

The bandgap energy of a semiconductor is an important parameter that determines its transport and optical properties as well as many other phenomena. The major technological advantage of the III-nitride system is that by alloying GaN with InN and/or AlN the bandgap can be tuned in a controllable fashion. Fig. 1.1 gives the bandgap energy versus chemical bond length for various materials.

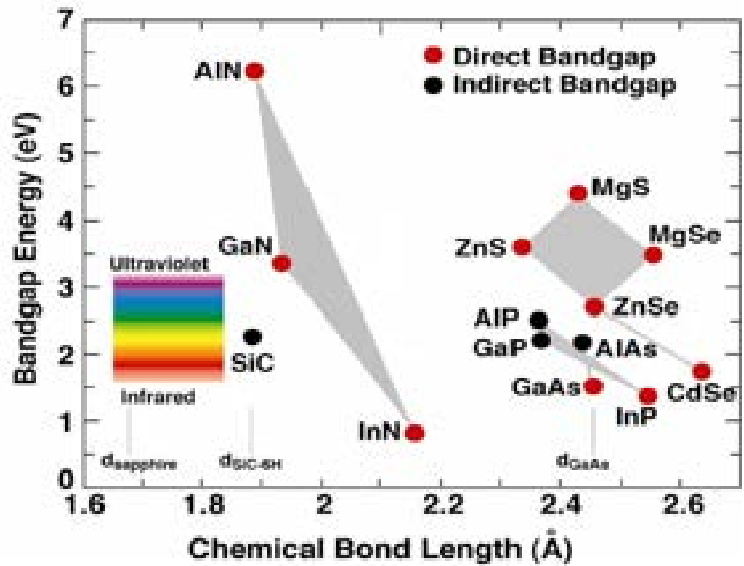

FIG. 1.1.   Bandgap energy versus chemical bond length for common semiconductors used for electronic and optoelectronic devices.[1]

In principle the bandgap energy of III-nitride alloys can be varied continuously from 0.7 eV (pure InN) to 6.2 eV (pure AlN) covering the spectral range from infrared to deep UV. In ternary compounds like InGaN, AlGaN, or InAlN the change of the bandgap energy with composition can be described by linear interpolation with the inclusion of a bowing parameter $b$. The bowing parameter represents the magnitude of the second order correction to the linear dependence. Some values of bowing parameters reported for the III-nitrides are given in Table 1.1. Note that in cases involving indium the bowing parameter can be dependent on the indium composition in the alloy. The bandgap energy of the III-nitride alloys is given by equation 1.1.

$$E_g(x) = xE_{g,AN} + (1-x)E_{g,BN} - bx(1-x) \qquad (1.1)$$

Where "A" and "B" represent Al, Ga, or In atoms, "$x$" is the composition of A ($0 \leq x \leq 1$) and "$b$" is the bowing parameter for the ABN alloy.

Table 1.1. Bowing parameters for III-nitrides.

| Alloy | Composition "$x$" | Bowing Parameter "$b$" |
|---|---|---|
| $Al_xGa_{1-x}N$ | $0 \leq x \leq 1$ | $1^2$ |
| | $0 \leq x < 0.45$ | $0.69^3$ |
| $In_xGa_{1-x}N$ | $0 \leq x \leq 0.5$ | $1.43^4$ |
| | $x \leq 0.12$ | $3.5^5$ |
| | $0 \leq x \leq 1$ | $2.77/(1+1.007x)^6$ |
| $In_xAl_{1-x}N$ | $0 \leq x \leq 0.85$ | $14.60/(1+4.53x)^6$ |
| | $0.25 \leq x$ | $3^7$ |

## 1.3. CRYSTAL STRUCTURE AND PHYSICAL PROPERTIES

III-nitride semiconductors normally crystallize in the stable wurtzite (hexagonal) structure, but under certain growth conditions they can exist as a zincblende (cubic) structure. The cubic and hexagonal phases differ only by the stacking sequence of close-packed III-N planes, and the energy difference between the two structures is small. Changing the sequence during growth produces defects such as stacking faults. Table 1.2 lists the energy bandgap and the lattice parameters for the two structures of III-nitrides ("w" stands for wurtzite and "z" for zincblende).

Table 1.2. Energy bandgap and lattice parameters for III-nitrides at 300K.[8, 9]

| Property | AlN | | GaN | | InN | |
|---|---|---|---|---|---|---|
| | w | z | w | z | w | z |
| Energy gap (eV) | 6.2 | 5.4 | 3.39 | 3.299 | 0.7 | 0.78 |
| Lattice constants (Å) | $^a$3.111 $^c$4.979 | 4.38 | $^a$3.189 $^c$5.185 | 4.50 | $^a$3.538 $^c$5.703 | 4.98 |

The zincblende structure belongs to the space group *F43m* and has an ABCABC stacking sequence of {111} close packed planes (Fig. 1.2), where the letter A, B or C represents a III-Nitrogen bond.

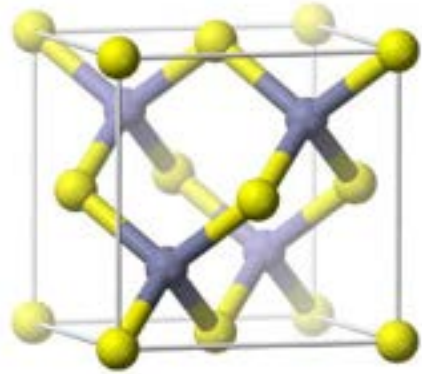

FIG. 1.2.   Schematic illustration showing the zincblende lattice structure of III-nitrides with N atoms represented by blue and group III-elements represented by yellow spheres.

The wurtzite structure has an ABABAB stacking sequence and belongs to the space group *P6_3mc*. This structure has alternating layers of close packed (0001) metal atoms (In, Ga, Al) and nitrogen atoms (Fig. 1.3).

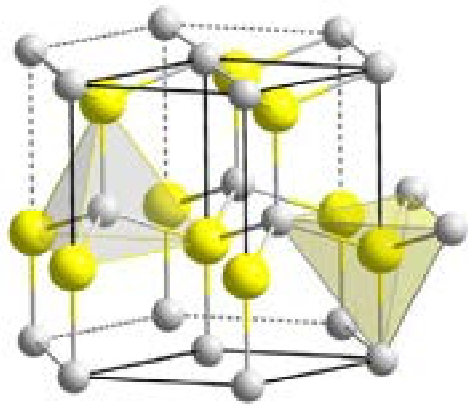

FIG. 1.3.   Schematic illustration showing the wurtzite lattice structure of III-nitrides with N atoms represented by gray and group III-elements represented by yellow spheres.

Each anion is surrounded by four cations at the corners of a tetrahedron, and vice versa.  The lack of a center of symmetry in the wurtzite structure of the III-nitride materials leads to spontaneous polarization. Also, the introduction of an external strain on this structure produces piezoelectric polarization.

The strong chemical bonding within III-nitrides results in high melting points, mechanical strength and chemical stability of these materials. Their strong bonding makes them resistant to high-current electrical degradation and radiation damage that is present in the active regions of light emitting devices.[1] These materials also possess good thermal conductivity. III-nitride based devices can operate at high temperatures as well as in hostile environments. Table 1.3 lists some of the physical properties of AlN, GaN and InN.

Table 1.3. Physical properties of III-nitride materials.[8]

| Property | AlN | GaN | InN |
| --- | --- | --- | --- |
| Thermal expansion coefficient (300K) $a$ ( $\times 10^{-6}$ K$^{-1}$) $c$ ( $\times 10^{-6}$ K$^{-1}$) | 4.2 5.3 | 5.59 3.17 | 5.7 3.7 |
| Melting point (°C) | >3000 | >2500 | >1100 |
| Cohesive energy per bond (eV) | 2.88 | 2.24 | 1.93 |
| Bond length (Å) | 1.89 | 1.94 | 2.15 |
| Thermal conductivity $\kappa$ (W/cm-K) | 2.0 | 1.3 | 0.8 |
| Bulk modulus (GPa) (300K) | 210 | 210 ± 10 | 140 |
| Refractive index, n | 2.2 | 2.35 | 2.56 |

## 1.4. THE DEVELOPMENT OF III-NITRIDES THROUGH THE YEARS

The synthesis of III-nitrides started in the early years of the 20th century[10-12], but their growth proved to be difficult. As a result there was no significant advancement in the growth of these compounds until the year 1969, when the growth of a GaN epilayer was reported by hydride vapor phase epitaxy (HVPE).[13] HVPE is a suitable growth method for producing bulk materials which can be used to study many of their physical properties. Metalorganic chemical vapor deposition (MOCVD) was first developed for the growth of GaN in 1971.[14] Thereafter, both HVPE[15, 16] and MOCVD[17, 18] techniques were used to grow GaN directly on sapphire, the latter being the most common technique today for producing epitaxial thin films.

Sapphire has been widely used as a substrate for the growth of III-nitride epilayers although it has a large lattice mismatch (~ 16% with GaN). This, combined with a difference in thermal expansion coefficient between GaN and sapphire substrates, resulted in poor GaN crystalline quality. The crystalline quality was improved when Amano *et al.*[19] used a two-step MOCVD growth technique, in which an A1N buffer layer was first deposited on sapphire prior to the growth of thick GaN layer. The thin A1N buffer layer accommodated the lattice mismatch and reduced the strain between GaN and sapphire. Another breakthrough was the demonstration of p-type conductivity of Mg-doped GaN in 1989.[20] Magnesium was used as a dopant and activated as an acceptor by low energy electron-beam irradiation (LEEBI). Nakamura *et al.* later on showed that Mg-doped GaN can also be activated by rapid thermal annealing at temperatures above 700°C in a $N_2$ atmosphere.[21] The use of thin GaN buffer layers instead of AlN buffer layers was shown to improve both crystalline quality and p-type conductivity of GaN layers.[22, 23] These advancements led to the fabrication of GaN-based blue light emitting diodes (LEDs) in 1991.[24]

Applications of the III-nitrides in the visible light regime require lowering the bandgap of GaN (3.4 eV), usually through the growth of InGaN layers. Following the success of growing GaN layers on sapphire, trial growths of InGaN layers directly on sapphire substrates gave unsatisfactory crystalline quality.[25, 26] The first high-quality InGaN layer was achieved on thick GaN buffer layer grown on sapphire.[27] Such layers were found to have strong band-to-band emissions at various indium compositions. The ability to grow high-quality InGaN layers with strong emissions has led to applications such as blue/green InGaN single quantum well structure LEDs,[28] blue AlGaN/InGaN double hetero-structure LEDs,[29] and room-temperature violet laser light emission using InGaN /GaN/ AlGaN hetero-structures under pulsed-operations.[30] In 1997 the development of the laser diode using InGaN multiple quantum well (MQW) structures was accomplished.[31]

The ultimate solution for the manufacture of high quality III-nitride materials is the growth of GaN or AlN bulk substrates for homoepitaxy of III-nitrides. In 2001, TDI Inc. announced fabrication of the first true bulk high quality GaN single crystal material. Bulk AlN substrates with the defect density as low as $10^{-3}$ $cm^{-2}$ have also been produced.[32] In July 2002, Sumitomo Electric Industries announced the development of a growth technique that allows reducing the density of dislocations in GaN to ~$10^6$ /$cm^2$. The Dislocation Elimination by Epitaxial growth with inverse-pyramidal Pits (DEEP) technique allows reducing dislocations by forming inverse-pyramidal pits on the surface of the grown material. GaN substrates 2-inch in diameter are now available for blue-violet lasers. Presently, Blu-Ray (from Sony) and HD-DVD (from Toshiba and NEC) drives are equipped with blue-violet GaN laser diodes operating at 405 nm.

Around the year 2000 the research field of non-polar III-nitrides started to emerge in an effort to eliminate the quantum-confined Stark effect (QCSE) in the quantum well (QW) structure, which in turn would lead to higher internal quantum efficiency. Waltereit *et al*[33] were able to grow m-plane GaN by molecular beam epitaxy and showed that m-plane GaN had a reduced radiative lifetime compared to the c-plane GaN. Since the internal quantum efficiency is inversely proportional to the radiative lifetime, m-plane GaN would open the way for high efficiency devices. Polarized light emission was confirmed from the *m*-plane LEDs by Lumileds in 2005.[34] Nonpolar m-plane InGaN/GaN laser diodes were realized in early 2007.[35]

The first *a*-plane GaN grown on *r*-plane sapphire was reported in 2002[36] and shortly thereafter a UV LED based on nonpolar *a*-plane GaN/AlGaN MQWs grown on *r*-plane sapphire was reported by Chen *et al*.[37] They demonstrated that the peak emission intensity was almost 30 times stronger than that in structures grown on *c*-plane sapphire. No shift of emission peak position was observed for nonpolar MQWs with increasing pump current or optical pump density, whereas devices built on *c*-plane structure exhibited a blue shift as large as 250 meV. In 2004, visible LEDs based on *a*-plane GaN/InGaN MQWs and emitting in the range of 413-425 nm were reported.[38, 39]

These advances led to the impressive research and development work in III-nitride based science and technology during the last decade. Although remarkable progress has been made in this field, much research is still needed to improve the efficiency and device performance of the III-nitride based devices.

## 1.5. CHALLENGES IN ACHIEVING HIGH EFFICIENCY III – NITRIDE LEDs

The efficiencies of III-nitride LEDs fall dramatically at wavelengths in the deep ultraviolet and in the green. Some of the current challenges related to III-nitride materials and their growth are: substrates used for GaN epitaxial growth, growth and phase stability of InGaN layers, lattice mismatch between InGaN/GaN, and the presence of piezoelectric fields due to lattice misfit strain.

### 1.5.1. SUBSTRATES FOR GaN GROWTH

Obtaining a suitable substrate for the growth of III-nitride epitaxial layers is one of the major issues encountered in the development of nitride-based devices. Despite considerable improvement in the growth of III-nitrides in the recent years, native substrates are still not commercially available due to their small size and high price. This results in the use of foreign substrates for their epitaxial growth. To date, sapphire is the most commonly used substrate for epitaxial growth of nitride materials. As explained in the previous section, the lattice mismatch and the difference in thermal expansion coefficients between the substrate and the film generate strain within the III-nitride epilayers, which in turn induces large electric fields. This yields to the quantum-confined Stark effect (QCSE), which decreases the radiative recombination rate and therefore the efficiency of LEDs. Mismatch of thermal expansion coefficients induces thermal stress in the film and substrate during cooling down to room temperature, which can lead to bowing and even cracking of the epitaxial films.

Lattice mismatch between the substrate and the epilayer also leads to high dislocation density in the order of $10^{10}$ cm$^{-2}$.[40] Fig. 1.4 shows a cross section transmission electron microscope (TEM) image of an LED thin film structure with a high dislocation density in the GaN film grown on sapphire substrate. The quality of films is improved with the addition of a buffer layer, which increases wetting of the III-nitride film and leads to growth of smooth films, resulting in better film quality.[19, 22]

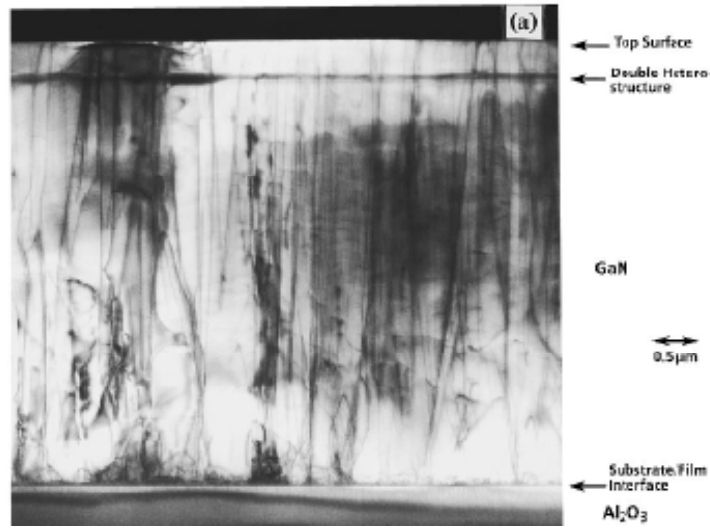

FIG. 1.4.  Cross-section TEM image of a blue LED, with an InGaN double heterostructure grown on a thick GaN layer by MOCVD on a sapphire substrate.[40]

Contrary to traditional III-V compounds where dislocation densities above ~$10^3$ cm$^{-2}$ lead to significantly reduced performance, the efficiency of InGaN based LEDs operating in the violet and blue wavelengths appear to be unaffected by such high dislocation densities.[1, 40] However, the device efficiency drops rapidly with increasing In composition for LEDs emitting in the green. Therefore, indium incorporation during growth, the compositional stability of InGaN thin films, and the large lattice mismatch of InGaN with GaN are important materials aspects to understand and overcome.

### 1.5.2. GROWTH OF InGaN: PHASE STABILITY

One of the most challenging issues for growth of indium-rich InGaN for use in the green emission devices is compositional instability that leads to the formation of In-rich and/or Ga-rich regions in the film. This non-uniform distribution has been shown to affect the optical properties of the as-grown films by increasing the FWHM and reducing the peak intensity.[41] On the other hand phase separation of InGaN can also form localized states. It has been argued that this could contribute to the high efficiency of the InGaN-based LEDs, because these localized energy states could capture the carriers before they are captured by the non-radiative recombination centers such as dislocations. [42-45] There is no doubt that the role of phase separation in InGaN is one area that needs much fundamental research.

### 1.5.3. PIEZOELECTRIC FIELDS

As a result of the lack of a center of symmetry, the wurtzite structure of the III-nitride materials exhibits pronounced polarization effects. This leads to the existence of a polarization charge at the hetero-interfaces. Also, the introduction of an external strain on this structure produces piezoelectric polarization. The piezoelectric effect has two components. One arises from lattice mismatch strain and the other from the thermal expansion coefficient difference between the substrate and the epitaxial layers. The macroscopic polarization in the material comprising the active region of the quantum wells (QWs) gives rise to a net electric field. This field will induce a spatial separation of the wave functions of electron and hole in the well. Therefore, the overlap of the wave function decreases and the radiative recombination rate is reduced. This phenomenon is called quantum-confined Stark effect (QCSE).

**1.6. ORGANIZATION OF THE DISSERTATION**

This dissertation constitutes a thorough study of the growth and thermodynamics of InGaN/GaN free-standing materials. A deep understanding of phase separation of InGaN should enable the formulation of new ideas for the production of more efficient devices based on these materials.

Chapter 2 focuses on phase separation and the thermodynamics of the InGaN system.

Chapter 3 describes the synthesis used for the growth of the freestanding InGaN samples. This chapter also gives a summary of the methods currently used for the growth of III-nitrides.

Chapter 4 details the characterization methods used to study the phase separation of the samples, namely x-ray diffraction (XRD), secondary electron microscopy (SEM), and cathodoluminescence (CL).

Chapter 5 presents characterization results of samples grown from Ga-In alloys and thermodynamics considerations of the growth.

Chapter 6 includes a study of the growth and characterization of GaN films grown at relatively low temperatures and showing the coexistence of cubic and hexagonal phases.

Chapter 7 presents characterization results of the InGaN powders grown by ammonolysis of a precursor.

Chapter 8 summarizes the importance of this work and gives suggestions for future research.

# CHAPTER 2

# PHASE SEPARATION

## 2.1. PHASE SEPARATION OF III-NITRIDES

A significant problem for the application of III-nitrides in electronic and optoelectronic devices is their tendency toward phase separation[46-55], which can unfavorably affect the alloy electronic and optical properties.[56] Therefore, the study of the growth and thermodynamics of the InGaN system has great scientific and commercial importance. The growth of high quality InGaN films, especially with compositions above 20% indium, is difficult because of phase separation. High temperatures are desirable for growth of nitride films in order to dissociate ammonia ($NH_3$) and free up nitrogen for bonding. However, the InN bond is significantly weaker than that for GaN at high temperatures, and InGaN films will tend to have difficulty with indium incorporation. If the temperature is lowered, hydrogen will etch nitrogen away from the indium leaving indium droplets on the growth surface. Indium incorporation, and an inherent miscibility gap in the InGaN alloys, can lead to compositional variations and spinodal phase separation over much of the compositional range. Also, large strain energies can be produced in InGaN films grown on GaN, due to the large difference in lattice parameter between GaN and InN. But lattice misfit relaxation is hindered by crystal geometry, where the basal slip planes lie perpendicular to the [0001] preferred crystal growth direction. Thus, when strain gets too high, phase separation can lead to a significant reduction of the InGaN strain energy.

Phase separation in the InN–GaN alloy system occurs through spinodal decomposition, where a miscibility gap develops between the two binary species, InN and GaN. Fig. 2.1 shows the free energy curve for a binary system and the regions on the phase diagram that are associated with spinodal decomposition.

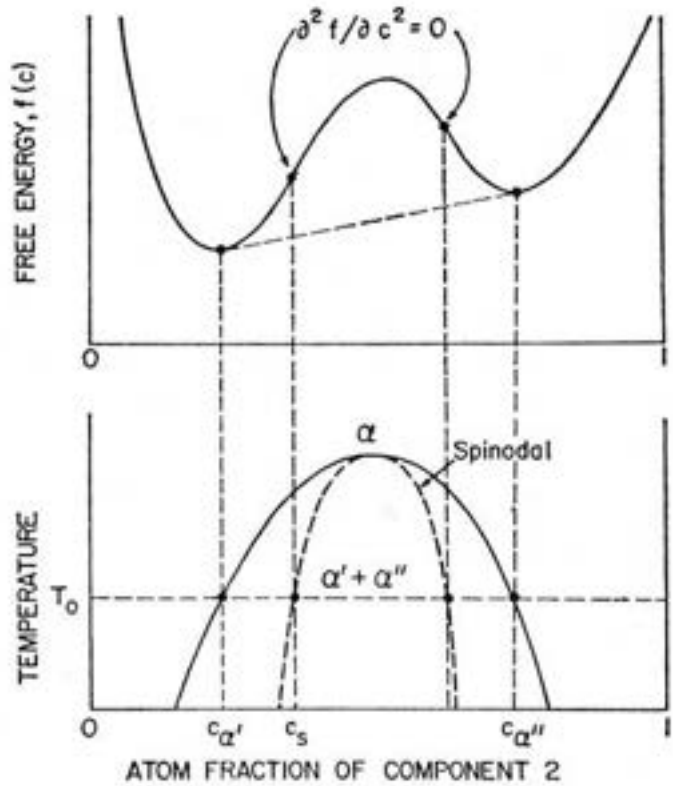

FIG. 2.1.  Free energy curve and the regions in the phase diagram associated with spinodal decomposition.[57]

The free energy $f$ of a binary system is plotted in Fig. 2.1 as a function of composition $c$ for a temperature below the critical temperature. Equilibrium phase compositions are those corresponding to the free energy minima. The boundary of the unstable region, sometimes referred to as the binodal or coexistence curve, is found by performing a common tangent construction of the free-energy diagram.

The spinodal decomposition regions are defined as the regions that have negative curvature in the corresponding free energy versus composition curve. Inside the spinodal region the free energy change is negative for small fluctuations in composition. The system is thus unstable and subject to composition fluctuations. As a result phase separation will occur. A more detailed overview of the phase diagrams of a binary alloy system including InGaN (pseudo-binary) is given in the next section.

Phase separation in InGaN was first observed during growth of polycrystalline samples after long annealing at temperatures above 600°C.[58] Ho and Stringfellow[59] calculated a theoretical phase diagram for the InN–GaN alloy system (Fig. 2.2) using a modified valence force field model. This calculated phase diagram shows that the equilibrium solubility of InN in bulk GaN is approximately 6% at typical growth temperatures of 700 to 800 °C. Phase separation through spinodal decomposition should occur in InGaN alloys containing more than ~20% indium at 800°C. These results were supported experimentally by El-Masry *et al.*[48] and by Ponce *et al.*[54] X-ray diffraction (XRD) analysis and TEM indicated phase separation in $In_xGa_{1-x}N$ layers with over 28% In composition grown by MOCVD. However, the calculations of Ho and Stringfellow[59] are done for relaxed InGaN material at equilibrium and as a result their phase diagram may not apply to strained InGaN layers.

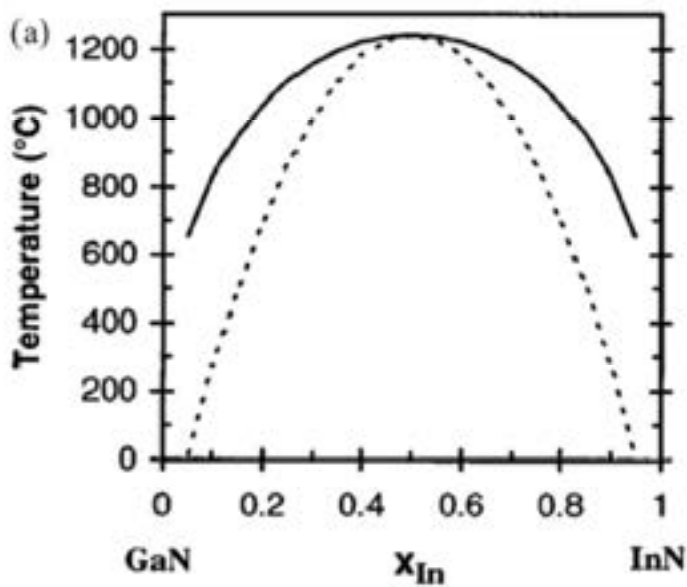

FIG. 2.2.  Phase diagram calculated by Ho and Stringfellow for InGaN.[59]
Binodal (solid) and spinodal (dashed) curves are calculated assuming a
constant average value for the interaction parameter.

The calculations of Karpov[60] for the phase diagram of biaxially strained InGaN produced a spinode with reduced miscibility gap and shifted toward higher indium contents than those predicted by Ho and Stringfellow.  The Karpov phase diagram is displayed in Fig. 2.3. More recent work by Liu and Zunger[61] also shows that phase separation may be suppressed by epitaxial strain. Some experimental support for the suppression of decomposition by epitaxial strain is provided by the work of Rao *et al.* [62] who studied InGaN epilayers more than 200 nm thick in TEM, and observed phase separation in relaxed regions of their films. However, regions of the films close to the InGaN/GaN interface did not show any evidence of phase separation.

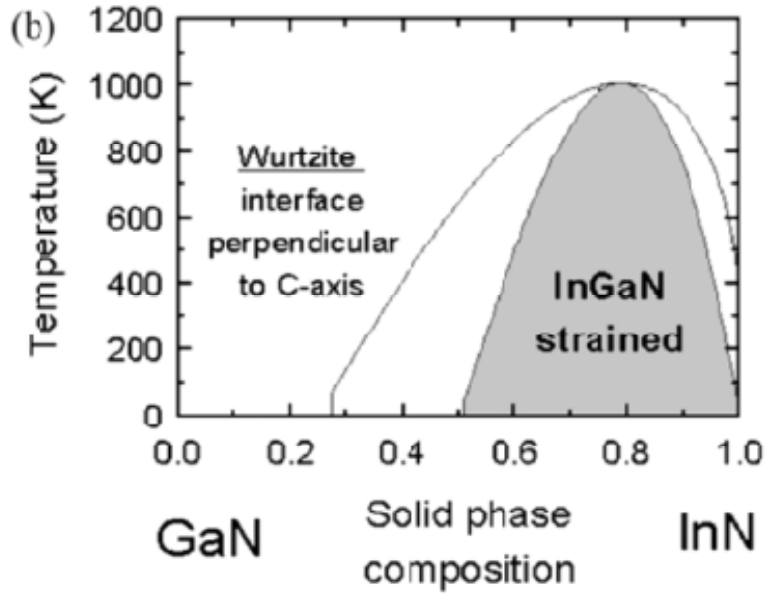

FIG. 2.3. Phase diagram calculated by Karpov for InGaN including the effects of biaxial strains present in epitaxial growth.[60]

Tabata et al.[63] also observed phase separation suppression in $In_xGa_{1-x}N$ alloy layers due to external biaxial strain, while in the relaxed $In_xGa_{1-x}N$ samples phase separation did occur. It is also suggested that the *composition pulling effect*[62,64] plays an important role in suppressing phase separation around $In_xGa_{1-x}N$/GaN interfaces. The composition pulling effect limits indium incorporation around interfaces due to the existence of large coherent strain. The lowered indium incorporation not only reduces the driving force for phase separation but also increases the critical thickness for relaxation. Upon growth, the layer progressively relaxes with more In incorporation, eventually triggering phase separation.

Even though extensive research has been done in studying the InGaN material, most of the existing research is on strained films. No experimental data exist for freestanding InGaN material, which instigated the research that led to this work.

## 2.2. THERMODYNAMICS OF BINARY ALLOYS

InGaN ternary solid solutions have important present or potential applications in light emitting devices such as LEDs and lasers. As a result, knowledge of the basic structural properties, thermodynamic stability and transport properties of these alloys is important. Growth techniques such as metal organic chemical vapor deposition (MOCVD) and molecular beam epitaxy (MBE) demand a deep knowledge of the alloy phase diagrams. The thermodynamic quantity needed to build phase diagrams is the molar Gibbs free energy of mixing, which is in turn connected with the microscopic structure of the alloy.

### 2.2.1. Gibbs' Free Energy and Equilibrium Criteria

The most convenient way to measure energy in phase systems is the Gibbs free energy. The Gibbs free energy is normally written as:

$$G = H - TS \qquad (2.1)$$

where $T$ is the temperature, $S$ is the entropy, and $H$ is the enthalpy of the system given by:

$$H = U + pV \qquad (2.2)$$

In the above equation U is the internal energy of the system, p is pressure, and V volume.

The second law of thermodynamics gives the fundamental criterion for equilibrium.

$$dS \geq \frac{dQ}{T} \tag{2.3}$$

We can rewrite the second law in the following form:

$$dQ - TdS \leq 0 \tag{2.4}$$

From the first law of thermodynamics with only pV work we have

$$dU = dQ - pdV \tag{2.5}$$

By combining (2.1), (2.2) and (2.5) the change in Gibbs free energy is:

$$dG = dH - TdS - SdT =$$
$$= dU + pdV + Vdp - TdS - SdT = \tag{2.6}$$
$$= dQ - TdS + Vdp - SdT$$

For a process at constant temperature and pressure equations (2.4) and (2.6) give

$$dG = dQ - TdS \leq 0 \tag{2.7}$$

It can be seen from equation (2.7) that at constant temperature and pressure the stable state of a system is the one that has the minimum value of the Gibbs free energy. Any spontaneous process in a system at constant T and p must decrease the Gibbs free energy (if the system is away from equilibrium) or leave the Gibbs free energy unchanged (if the system is at equilibrium). Thus, the phase stability in a system can be determined from knowledge of the variations of the Gibbs free energies with composition and temperature of the various possible phases.

## 2.2.2. Gibbs Free Energy of Binary Systems

**A. Ideal solutions**

In an isobaric and isothermal process of mixing of substances A and B the change in Gibbs free energy as a result of mixing is:

$$dG_{mix} = dH_{mix} - TdS_{mix} \qquad (2.8)$$

This $dG_{mix}$ is the difference between the free energy of the solution/mixture and the free energy of the pure components A and B. The enthalpy term, $dH_{mix}$, represents the nature of the chemical bonding, or, in different words, the extent to which A prefers B, or A prefers A as a neighbor. The entropy term, $dS_{mix}$, signifies the increase in disorder in the system as we let the A and B atoms mix. It is independent of the nature of the chemical bonding.

An ideal solution is one in which the atoms are randomly mixed and the interactions between species are identical, in other words the bond energy doesn't change whether the bond is A-A, B-B, or A-B. Therefore, for an ideal solution there is no overall enthalpy change. Then the Gibbs free energy of mixing becomes:

$$dG_{mix} = -TdS_{mix} \qquad (2.9)$$

Now let's assume we have a crystal with a total of N sites available for the occupation of atoms or molecules, n of which are occupied by A atoms/molecules and (N - n) are occupied by B atoms/molecules. In this case it can be shown that the total number of ways of distributing A and B is given by:

$$W = \frac{N!}{n!\,(N-n)!} \qquad (2.10)$$

The entropy of mixing components A and B is given by Boltzman's equation:

$$dS_{mix} = k_B \ln W \qquad (2.11)$$

where $k_B$ is Boltzmann's constant and $W$ is the number of the different arrangements of the atoms on lattice sites of the crystal given by equation (2.10). Replacing $W$ in equation (2.11) gives:

$$dS_{mix} = k_B \ln\left(\frac{N!}{n!\,(N-n)!}\right) \qquad (2.12)$$

Using Stirling's approximation $N \ln N! = N \ln N - N$ for large $N$, equation (2.12) becomes

$$dS_{mix} = k_B[N \ln N - n \ln n - (N-n)\ln(N-n)] =$$

$$= -k_B\left[n \ln\left(\frac{n}{N}\right) + (N-n)\ln\left(\frac{N-n}{N}\right)\right] =$$

$$= -Nk_B\left[\frac{n}{N}\ln\left(\frac{n}{N}\right) + \frac{(N-n)}{N}\ln\left(\frac{N-n}{N}\right)\right] \qquad (2.13)$$

If we consider $N$ to be Avogadro's number, $N_A = 6.02 \times 10^{23}$ atoms or molecules, then $N k_B = R = 8.314$ J/mole-K. The mole fractions of components $A$ and $B$ are given by $x_A = n/N$ and $x_B = (N-n)/N$ respectively, so equation (2.13) reduces to:

$$dS_{mix} = -R(x_A \ln x_A + x_B \ln x_B) \qquad (2.14)$$

This defines the ideal entropy change of mixing. Then the Gibbs free energy of mixing for an ideal solution is given by:

$$dG_{mix} = RT(x_A \ln x_A + x_B \ln x_B) \qquad (2.15)$$

It is clear that in the case of the ideal mixing of two components the change in free energy $dG_{mix}$ is always negative, since $x_A$ and $x_B$ are always less then unity.

**B. Regular Solutions**

The regular solution model is the simplest model used to describe the thermodynamic properties of semiconductors. A regular solution is a solution that diverges from the behavior of an ideal solution only moderately. The components mix in a completely random manner just like in the ideal solution, so the entropy is still the same as in the ideal mixing. The additional assumptions pertaining to the regular solution model are: (a) interaction between the constituent atoms occurs only between nearest neighbor pairs, and (b) the atoms reside on a lattice with each atom surrounded by Z neighbors.[65] By summing nearest-neighbor bond energies we get an enthalpy of mixing dependent on the composition of the components $dH_{mix} = \Omega x_A x_B$, where $\Omega$ is the interaction parameter term given by $\Omega = ZN_A [E_{AB} - \frac{1}{2}(E_{AA} + E_{BB})]$.[65] Then the Gibbs free energy of mixing for the regular solution model is:

$$dG_{mix} = \Omega x_A x_B + RT(x_A \ln x_A + x_B \ln x_B) \qquad (2.16)$$

From the Gibbs free energy curve, it is possible to generate the phase diagram for a given system. In the case where the entropy term is large compared to the enthalpy term, or the enthalpy term is negative, the free energy has only one minimum and in that case there is complete miscibility of the components. This is shown in Fig. 2.4, where the Gibbs free energy is calculated using equation (2.16) for $\Omega$ = 20 kJ/mol and T = 2000 K.

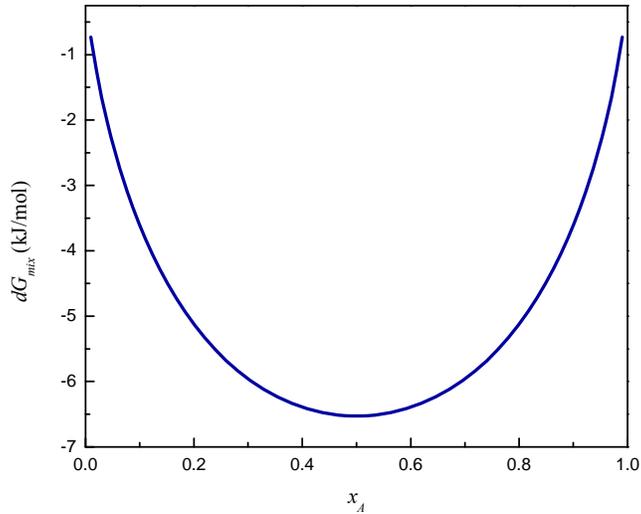

FIG. 2.4. The Gibbs free energy for a binary alloy when there is complete miscibility of the components calculated from equation (2.16) for $\Omega = 20$ kJ/mol and $T = 2000$ K.

In the case where the enthalpy is positive but comparable to the entropy term at a given temperature, the phase diagram is determined by the shape of the free energy curve as the one shown in Fig. 2.5, which is calculated from equation (2.16) for $\Omega = 20$ kJ/mol and $T = 850$ K.

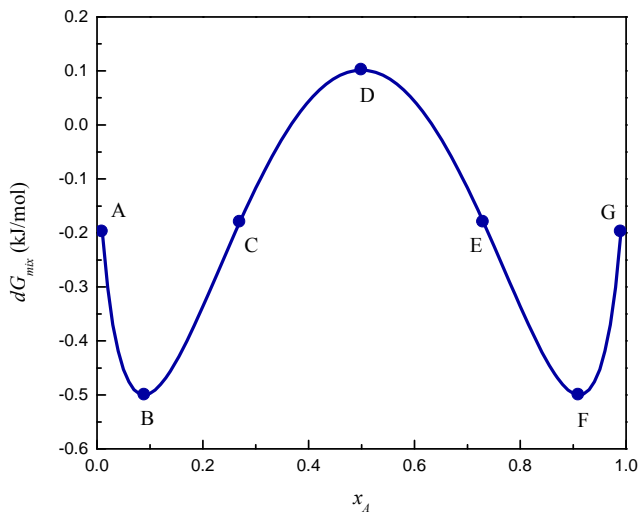

FIG. 2.5. The Gibbs free energy for a binary alloy when the enthalpy of mixing is larger than the entropy term calculated from equation (2.16) for $\Omega = 20$ kJ/mol and $T = 850$ K.

The Gibbs free energy curve in Fig. 2.5 shows two minima (points B and F) and a maximum (point D). The minima represent the equilibrium compositions. Any mixture with composition between points B and F will separate into two phases with equilibrium compositions B and F.

Any composition within the region between the two equilibrium points is not considered to be at equilibrium. Near the equilibrium points the compositions are not at equilibrium and metastable. However, if you move too far from the equilibrium you can enter the spinodal region, where the mixture is unstable. The spinodal region is defined where the free energy curve has an inflection point (points C and E).

Within the spinodal region, any microscopic fluctuations (which are impossible to prevent) are sufficient to set up a chain reaction decomposing the composition into its equilibrium phases. Both the equilibrium compositions (binodal) and the spinodal region are defined by the free energy vs. composition curve. The binodal points $(x,T)$ are those for which the first derivative of the Gibbs free energy of mixing is zero (minima). The spinodal points $(x,T)$ are those for which the second derivative of the Gibbs free energy of mixing is zero (inflection points).

Fig. 2.6 shows the molar enthalpy, entropy, and Gibbs free energy of mixing of two components when the enthalpy of mixing is positive and dominates the entropy term, calculated for for $\Omega = 20$ *kJ/mol* and $T = 400$ *K*.

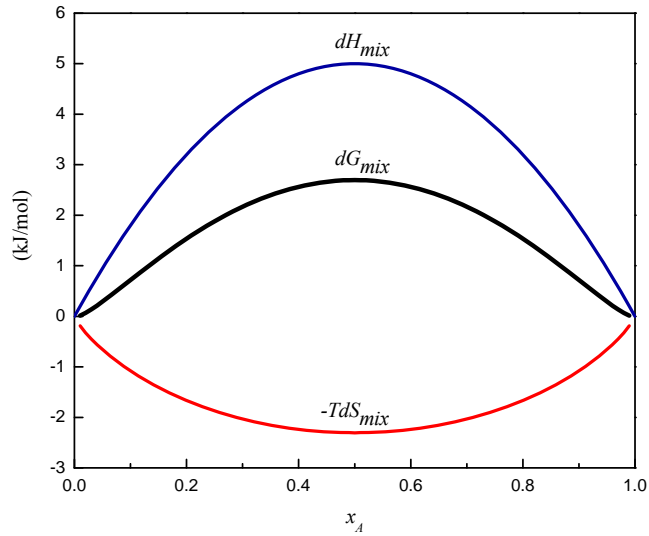

FIG. 2.6.  The enthalpy, entropy, and Gibbs free energy of mixing of a binary alloy for large positive enthalpy calculated for for $\Omega = 20$ kJ/mol and $T = 400$ K.

The Gibbs free energy of mixing is symmetrical about $x = 0.5$, at which point the entropy of mixing term produces a minimum and the enthalpy term produces a maximum.  As can be seen, $dG_{mix}$ is positive for almost all compositions making the binary system unstable at that temperature and completely immiscible.

Fig. 2.7 gives three calculated $dG_{mix}$ curves given by equation (2.16) for different temperatures.

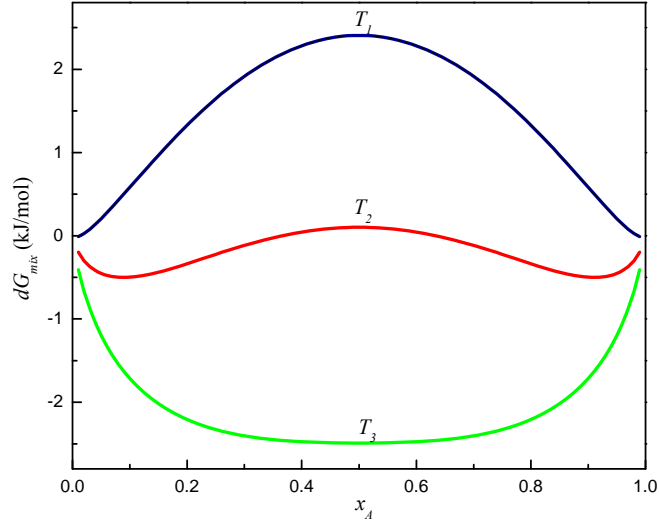

FIG. 2.7.  Gibbs free energy of mixing for a regular solution calculated from equation (2.16) for different temperatures $T_1 < T_2 < T_3$ and $\Omega = 20$ kJ/mol.

At low temperature $T_1 = 450\ K$ the $dG_{mix}$ curve has only a maximum and the system is completely immiscible. As the temperature increases to $T_2 = 850\ K$ the $dG_{mix}$ curve has two minima and a maximum in between, which results in phase separation.  At high temperatures $T_3 = 1300\ K$ the entropy dominates and the system is fully miscible with $dG_{mix}$ having only one minimum. The minimum value of $T$ for which the system becomes fully miscible is called the critical temperature of solubility.

## 2.3. PHASE DIAGRAM OF RELAXED InGaN SYSTEM

The regular solution model can be applied to $In_xGa_{1-x}N$ ternary alloy. The composition of [In] is represented by the variable $x$ and that of [Ga] by $(1 - x)$, so the Gibbs free energy of mixing in equation (2.16) takes the form:

$$dG_{mix} = \Omega x(1-x) + RT(x \ln x + (1-x)\ln(1-x)) \qquad (2.17)$$

The second term in the above equation is the ideal configuration entropy of mixing based on the assumption of a random distribution of Ga and In atoms. The interaction parameter in equation (2.17) is found using the valence force field (VFF) model. The mixing enthalpy is considered to be equal to the strain energy caused by the difference of the equilibrium bond lengths between the ternary alloy and the binary systems that compose the alloy. The VFF model treats both the stretching and bending of the bonds without adjustable parameters. In the case of InGaN ternary alloy, the lattice is composed of five types of tetrahedra, where one N atom resides at the center and In/Ga atoms are at the apexes of each tetrahedron. Following a similar approach as the one in Ref. 59, the strain energy ($E_m$) can be written:

$$E_m = \frac{3}{8}\sum_{i=1}^{4} \alpha_i \frac{(d_i^2 - d_{i0}^2)^2}{d_{i0}^2} + \frac{6}{8}\sum_{i=1}^{4}\sum_{j=i+1}^{4} \frac{(\beta_i + \beta_j)}{2}\frac{(d_i \cdot d_j + d_{i0}d_{j0}/3)^2}{d_{i0}d_{j0}} \quad (2.18)$$

where $d_i$ is the distance between the central atom *N* and a corner atom *In* or *Ga* in the tetrahedron, $d_{i0}$ is the equilibrium bond length in the binary compound GaN or InN (~4.52 Å for GaN and 4.98 Å for InN), $\alpha$ is the bond stretching force constant force (~81.09 N/m for GaN and 63.5 N/m for InN), and $\beta$ is the bond bending force constant (~12.16 N/m for GaN and 8.05 N/m for InN).[59] The enthalpy of mixing for the alloy is found from the summation of $E_m$ over all five possible tetrahedral arrangements which is weighted by the probability distribution. By fitting this mixing enthalpy to the regular solution form *Ωx(1-x)* the value of *Ω* is calculated to be *25.02 kJ/mol*.

The Gibbs free energy of mixing for the InGaN binary system given by equation (2.17) is calculated and graphed for different temperatures as shown in Fig. 2.8.

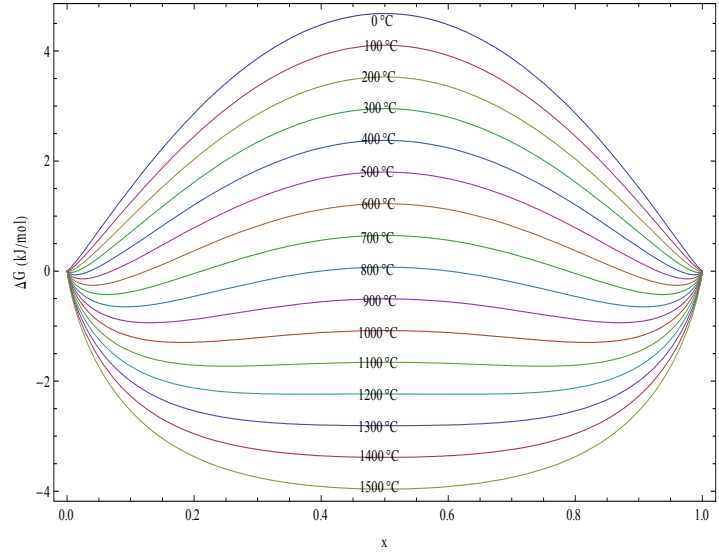

FIG. 2.8.  Gibbs free energy of mixing for InGaN calculated at different temperatures according to the regular solution model given by equation (2.17).

The phase diagram is constructed by finding the binodal and spinodal points for different temperatures. The binodal points $(x,T)$ are those for which the first derivative of the Gibbs free energy of mixing is zero (minima). The spinodal points $(x,T)$ are those for which the second derivative of the Gibbs free energy of mixing is zero (inflection points). Fig. 2.9 shows the calculated binodal (dashed line) and spinodal (solid line) curves for the InGaN alloy using the regular solution model with $\Omega = $ *25.02 kJ/mole.*

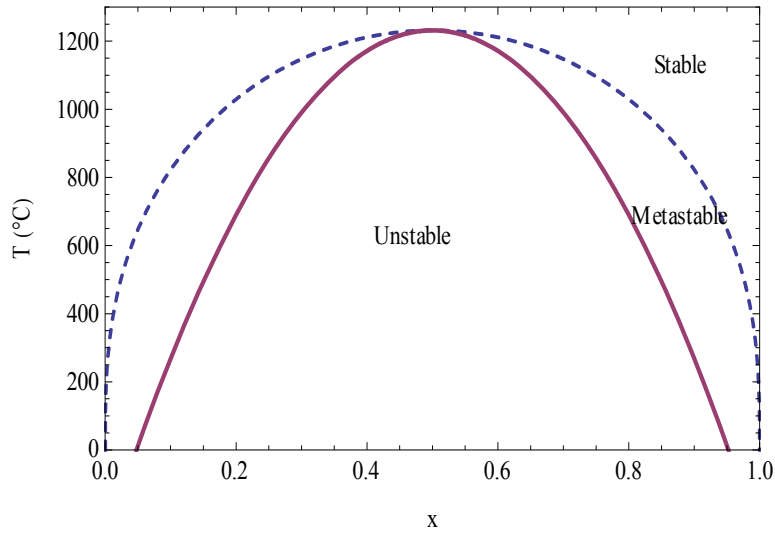

FIG. 2.9. Phase diagram of the binary alloy of InGaN. The dashed blue line and the solid red line represent the binodal and the spinodal curve respectively.

In the region outside the binodal curve only one phase exists and the system is completely miscible. The critical temperature is calculated to be 1236°C. The metastable region is between the binodal and the spinodal curves. Phase separation occurs inside the spinodal curve.

The validity of the application of the regular solution model to the InGaN system has been studied by Monte Carlo molecular simulation,[66] which found that the excess entropy is very small (zero for a regular solution), and the excess enthalpy is insensitive to temperature, indicating that the regular solution treatment is appropriate for this ternary alloy. The simulations also showed the critical solution temperature to be ~1550 K, in good agreement with the regular solution model used in this study (~1509 K).

## 2.4. CONCLUSIONS

The thermodynamics of phase separation for the InGaN ternary system has been analyzed using a strictly regular solution model. Graphs of Gibbs free energy of mixing were produced for a range of temperatures. Binodal and spinodal decomposition curves show the stable and unstable regions of the alloy in equilibrium. The interaction parameter used in the analysis is obtained from a strain energy calculation using the valence force field model. According to the calculated results, the critical temperature for InGaN is found to be 1236°C. This suggests that at typical growth temperatures of 800 – 1000°C a wide unstable two-phase region exists in InGaN. Phase separation in epitaxially grown InGaN has been experimentally reported to occur at an indium composition of more than 20%. The predicted temperature and composition region for phase separation are consistent with the experimentally observed difficulties in achieving high In content InGaN required for green emitting devices. These difficulties could be addressed by studying the growth and thermodynamics of these alloys. Knowledge of thermodynamic phase stabilities and of pressure – temperature – composition phase diagrams is important for an understanding of the boundary conditions of a variety of growth techniques. The above study applies only to growth processes under thermal equilibrium conditions and many techniques employed for the growth of InGaN use conditions away from equilibrium. Understanding phase diagrams and the degree of departure from the equilibrium state which corresponds to the so called 'driving force', is useful for these growth processes.

CHAPTER 3

III-NITRIDE GROWTH TECHNIQUES

Many growth techniques have been successfully developed for epitaxial growth of III-nitride thin films. Some of the most common techniques include: hydride vapor phase epitaxy (HVPE), metal organic chemical vapor deposition (MOCVD), and molecular beam epitaxy (MBE).

## 3.1. HVPE TECHNIQUE

HVPE was first developed for *GaN* growth by Maruska.[13] In the first part of this process, *HCl* vapor flows over hot *Ga* melt to form *GaCl₃* by the following chemical reaction:

$$2Ga + 2HCl \rightarrow 2GaCl_3 + H_2 \qquad (3.1)$$

Next, *GaCl₃* vapor is transported downstream to react with $NH_3$ producing GaN:

$$GaCl_3 + NH_3 \rightarrow GaN + HCl + H_2 \qquad (3.2)$$

The HVPE technique is comparatively simple and less expensive than other techniques used for growth of III-nitride. This technique has also a high growth rate (hundreds micron per hour), which is why it is now employed to grow thick GaN layers. For the growth of optoelectronic and electronic devices such as LEDs and high electron mobility transistors (HEMTs), accurate control of the film thickness and quality is crucial and hence either MOCVD or MBE are typically chosen rather than HVPE.

## 3.2. MOCVD GROWTH TECHNIQUE

The MOCVD technique, also known as organo-metalic vapor phase epitaxy (OMVPE), was developed in the late sixties by Manasevit,[14] who applied the technique to the growth of GaN on sapphire substrates. It is so far the most popular method for growing III-nitride films for commercial purposes. MOCVD, uses ammonia ($NH_3$) and metalorganic gases such as tri-methyl-X, $(CH_3)_3X$, with X being Ga, Al or In, as reactant sources for the growth of III-nitride films. Hydrogen carrier gas is used to transport the reactant gases to the surface of a rotating substrate located inside a chamber, which is typically kept at around 1000°C. The reaction of gases at the substrate yields a deposited film according to the chemical equation below:

$$(CH_3)_3Ga + NH_3 \rightarrow GaN + 3CH_4 \qquad (3.3)$$

A schematic diagram of an MOCVD system is shown in Fig. 3.1 below.

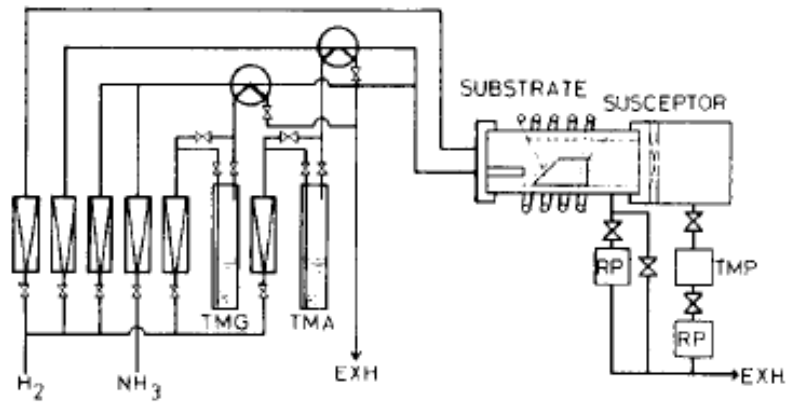

FIG. 3.1. Schematic MOCVD system for the growth of III – nitrides.[67]

Separate inlets are used for different reactants to minimize the pre-deposition reaction of the gases. Rotation of the substrate is designed to improve the uniformity of the film. The MOCVD growth technique provides excellent uniformity in layer thickness, composition and carrier concentration over a large area wafer.

The disadvantage of this method resides at the high substrate temperature necessary to decompose $NH_3$. In fact, due to the thermal mismatch with the substrate, postgrowth cooling introduces significant amounts of strain and structural defects into the nitride film. Residual contaminants can also be introduced in the film at these high temperatures and III-metal desorption, diffusion and segregation can occur.

### 3.3. MBE GROWTH TECHNIQUE

An alternative growth technique that allows reducing the substrate temperature is MBE. The MBE process is usually performed at relatively low temperatures of around 650°C – 800°C. Solid sources of *Ga*, *In*, *Al*, *Mg* and *Si* are typically used for the growth of the III-nitride alloys. Particularly, *Mg* and *Si* are used for p- and n-doping respectively. Nitrogen ($N_2$) and group III vapor atoms are directed toward a heated substrate and react according to the following chemical reaction:

$$2Ga + N_2 \rightarrow 2GaN \qquad (3.4)$$

To facilitate efficient growth of III-nitrides, the nitrogen species have to be activated. This is because molecular nitrogen is inert, due to the strong N-N bond. Radio frequency (RF) is a common technique used for the activation of nitrogen. Alternatively $NH_3$ can be used as the nitrogen source, as in the case of the ammonia-MBE system.

The MBE process is an ultrahigh-vacuum (UHV) thin film deposition technique. Such UHV conditions enable the growth of semiconductor epitaxial layers with high purity. In addition, due to the ultrahigh-vacuum environment for its growth process, a number of insitu monitoring techniques such as reflection high-energy electron diffraction (RHEED) and gas analysis process such as quadrupole mass spectrometry can be incorporated into the growth chamber. These insitu analytical methods greatly facilitate the characterization and optimization of the growth parameters. As a result, the MBE process has been transferred to production-scale manufacturing of optoelectronic and electronic devices. A schematic diagram of a typical MBE system is shown in Fig. 3.2.

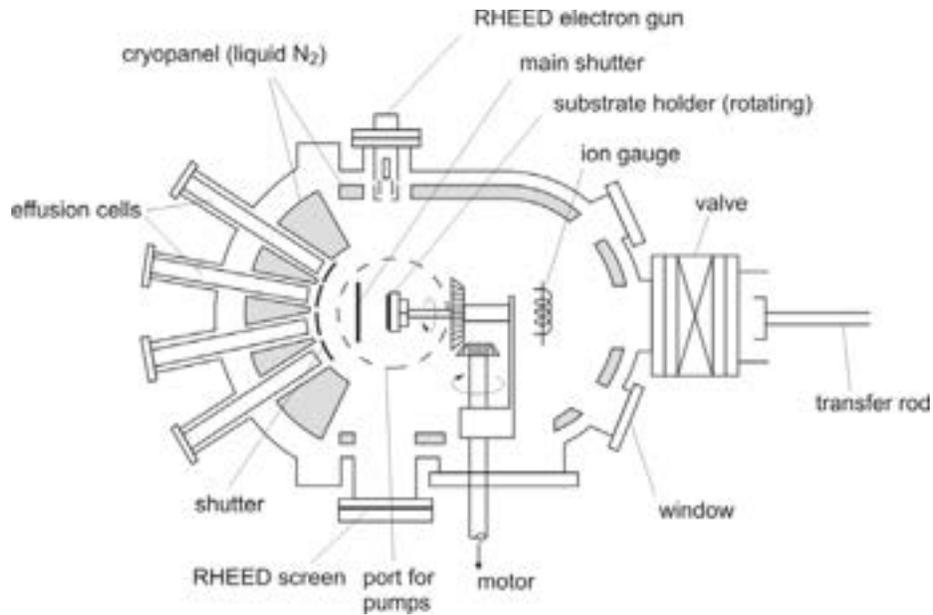

FIG. 3.2.   Schematic MBE system for the growth of III – nitrides.[68]

The main disadvantages of MBE for III-nitride growth are related to the growth rate. This rate is limited by the available flux of low kinetic energy of reactive nitrogen ions. While attempting to increase this rate, *Ga* droplets are observed to form on the surface. Moreover, when the power is increased, more energetic ions are created. These more energetic ions are in general detrimental to crystal quality since they can introduce point defects.

## 3.4. SYNTHESIS OF InGaN POWDERS

### 3.4.1. Synthesis of InGaN by ammonolysis of the Ga-In metal alloy

The procedure for the synthesis of InGaN from the Ga-In metal alloy consists of two basic steps. In the first step, the desired masses of ultrahigh-purity gallium and indium metals are placed in a high-alumina crucible. In order to avoid any contamination, the masses of gallium and indium are measured inside an autoclave filled with nitrogen. These masses are calculated based on the desired composition of indium in the final product of InGaN as given by equation (3.1).

$$m_{In} = \#moles_{In} * M_{In} \qquad (3.1)$$

In equation (3.1), $m_{In}$ is the mass of indium metal in *grams*, $\#moles_{In}$ is the number of moles of indium, and $M_{In}$ is the molar mass of elemental indium in *grams/mol*. The mass of gallium is calculated using the same equation, but with gallium as the element. Knowing the desired composition $x$ of indium in $In_xGa_{1-x}N$ and the number of moles of gallium, the number of moles of indium used in equation (3.1) is found according to equation (3.2) below:

$$x = \frac{\#moles_{In}}{\#moles_{In} + \#moles_{Ga}} \qquad (3.2)$$

The alumina crucible containing the Ga-In alloy is placed inside a stainless steel vessel that is fitted to a mechanical shaker (see FIG. 3.3).

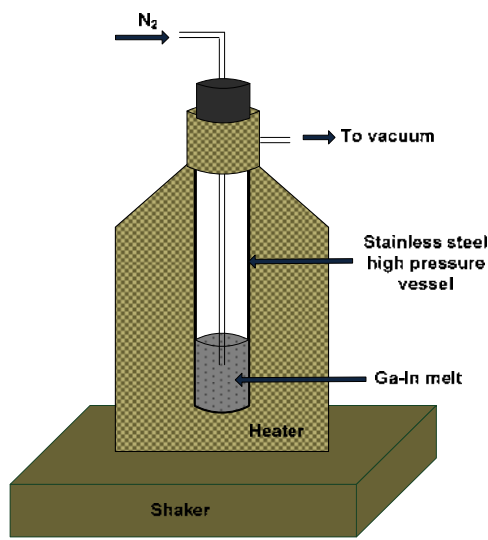

FIG. 3.3. The schematic of the mechanical shaker and heater to produce the Ga-In-$N_2$ solution.

The vessel is hermetically closed and is heated under vacuum ($10^{-3}$ Torr) from room temperature to ~250°C. Then, the vessel is filled with high-purity nitrogen and shaken for ~30 minutes in order to produce a highly homogeneous gas – liquid solution. The solution is then poured into an alumina boat.

The second step consists of the reaction of the prepared solution of Ga – In with ammonia inside a horizontal quartz tube reactor (see Fig. 3.4). The quartz tube reactor contains three zones (zone 1, zone 2, and zone 3) that can be heated at different temperatures. Initially, the prepared solution of Ga-In is placed at the entrance of the reactor. Then, the tube reactor is evacuated with a mechanical pump to ~$10^{-3}$ Torr, while simultaneously zone 2 is heated to ~1100°C and zone 3 to ~900°C. These are the optimum temperatures that produce a reaction. The alumina boat containing Ga-In solution is kept at the entrance of the tube (~200°C) while high-purity $N_2$ flows through the quartz tube. Zone 2 and zone 3 of the reactor reach the desired temperature after approximately 1 hour. At this time the vacuum system is closed and a current of ultrahigh-purity ammonia (~330 sccm, 760 Torr) flowed through the reactor. After steady-state conditions are reached (~30 min), the boat with the Ga-In solution is moved quickly to zone 2 of the reactor using a magnetic manipulator (see FIG. 3.4).

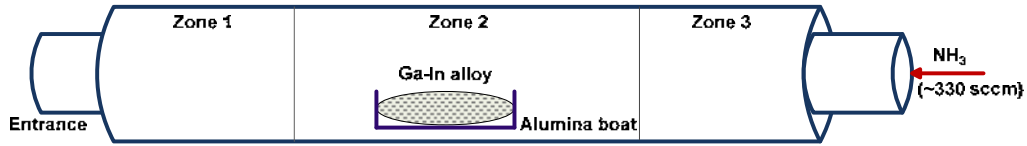

FIG. 3.4. A schematic of the quartz tube reactor showing the three zones and the Ga-In solution placed in zone 2.

The Ga-In melt reacts with ammonia according to the following reactions:

$$Ga(l) + NH_3(g) \rightarrow GaN(s) + \frac{3}{2}H_2(g) \quad (3.3)$$

$$In(l) + NH_3(g) \rightarrow InN(s) + \frac{3}{2}H_2(g) \quad (3.4)$$

After the reaction happens, the boat with the product is moved to the entrance of the reactor (room temperature) using the magnetic manipulator. Then the ammonia flow is closed and nitrogen is flown through the quartz tube. After the product cools, the boat is removed from the reactor and the powder is ground in a mortar and stored for analysis.

### 3.4.2. Synthesis of InGaN by ammonolysis of a complex salt

The procedure for the synthesis of $In_xGa_{1-x}N$ is the method used by Garcia et al.[69] with a few modifications. The first step in this procedure is the preparation of the precursor ammonium hexafluoroindium-gallate $[(NH_4)_3In_xGa_{1-x}F_6]$, which first requires the reaction of the aqueous solution of indium nitrate [$In(NO_3)_3$], gallium nitrate [$Ga(NO_3)_3$] and ammonium hydroxide ($NH_4OH$) to form indium-gallium hydroxide [$In_xGa_{1-x}(OH)_3$] according to the following chemical reaction:

$$xIn(NO_3)_3(aq) + (1-x)Ga(NO_3)_3(aq) + 3NH_4OH(aq) \rightarrow$$

$$\rightarrow In_xGa_{1-x}(OH)_3(s) + 3NH_4NO_3(aq) \qquad (3.5)$$

The required amount of each reactant is calculated from the desired concentration $x$ of indium and the desired mass of $In_xGa_{1-x}N$. The equations used for finding the exact mass of the reactants are given below.

$$m_{In(NO_3)_3} = x * M_{In(NO_3)_3} * \frac{m_{InGaN}}{M_{InGaN}} \qquad (3.6)$$

In equation (3.6) $m_{In(NO3)3}$ is the mass of solid indium nitrate in *grams*; $x$ is the desired concentration of indium in $In_xGa_{1-x}N$; $M_{In(NO3)3}$ is the molecular mass of indium nitrate in *grams/mol*; $m_{InGaN}$ is the mass of $In_xGa_{1-x}N$ in *grams*; and $M_{InGaN}$ is the molecular mass of $In_xGa_{1-x}N$ in *grams/mol*. Similarly, for the mass of solid $Ga(NO_3)_3$ we use the following equation:

$$m_{Ga(NO_3)_3} = (1-x) * M_{Ga(NO_3)_3} * \frac{m_{InGaN}}{M_{InGaN}} \qquad (3.7)$$

Ammonium hydroxide is in liquid form, so a volumetric amount is calculated using the equation below.

$$V_{NH_4OH} = 3 * \frac{M_{NH_4OH}}{D_{NH_4OH}} * \frac{m_{InGaN}}{M_{InGaN}} \qquad (3.8)$$

In equation (3.8) $V_{NH4OH}$ is the volume of ammonium hydroxide in *ml* and $D_{NH4OH}$ is the density of ammonium hydroxide in *grams/ml*.

The solid nitrates with masses calculated according to equations (3.6) and (3.7) are placed in a quartz beaker and dissolved in deionized water. Then, the appropriate volume of ammonium hydroxide calculated by equation (3.8) is added. The product formed is a white viscous solution constituting of a mixture of solid indium gallium hydroxide and ammonium nitrate ($NH_4NO_3$) according to reaction (3.5).

Indium gallium hydroxide is required to react with ammonium fluoride ($NH_4F$) in order to produce the precursor ammonium hexafluoroindium-gallate according to the following reaction:

$$In_xGa_{1-x}(OH)_3(s) + 6NH_4F(aq) \rightarrow (NH_4)_3In_xGa_{1-x}F_6(s) + 3H_2O(l) + 3NH_3(g) \quad (3.9)$$

To separate the solid compound indium gallium hydroxide from the ammonium nitrate salt, a filtration system is used. The white product is washed with deionized water until no change in the PH of the water solution coming through the filter is registered. As a result, indium gallium hydroxide, in the form of a thick white paste, is left on the filter. The cleaned indium gallium hydroxide is placed in a Teflon beaker and reacted with the ammonium fluoride concentrated solution in order to produce the precursor ammonium hexafluoroindium-gallate. The amount of ammonium fluoride used for this reaction is calculated according to the equation below:

$$m_{NH_4F} = 6 * M_{NH_4F} * \frac{m_{InGaN}}{M_{InGaN}} \quad (3.10)$$

The Teflon beaker is placed into a vacuum dry oven and heated at 140°C for 2 hours to dry the product (ammonium hexafluoroindium-gallate) by allowing for the evaporation of the water and escape of the ammonia gas. The procedure for the synthesis of the precursor $(NH_4)_3In_xGa_{1-x}F_6$ is illustrated in FIG. 3.5.

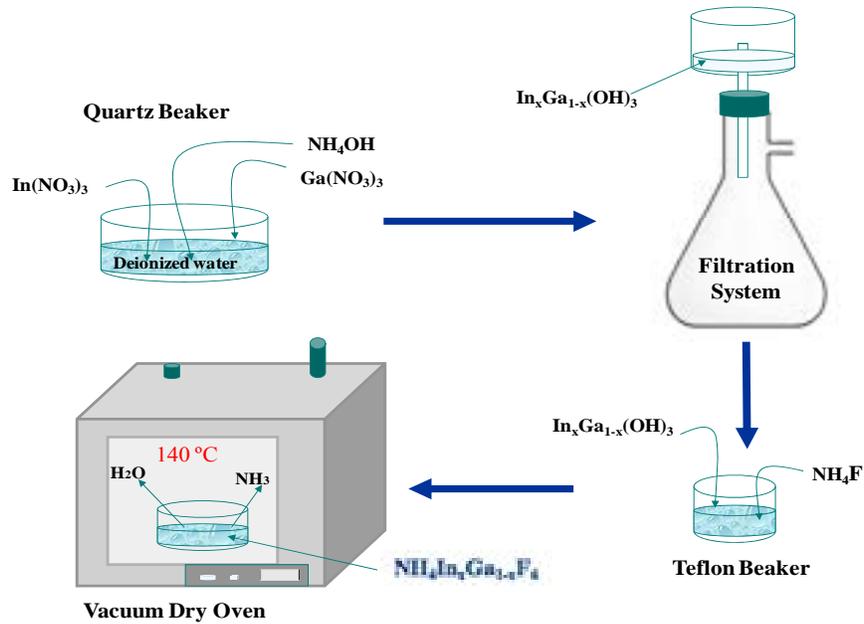

FIG. 3.5.   The procedure used for the synthesis of the precursor ammonium hexafluoroindium-gallate.

The second step in the synthesis of InGaN involves the ammonolysis of the ammonium hexafluoroindium-gallate in a tubular reactor.  The reactor consists of a tubular resistance heating furnace ~50 cm in length, inside which resides a cylindrical quartz tube ~2" in diameter and ~1m in length with stainless steel rims at both sides (see FIG. 3.6).

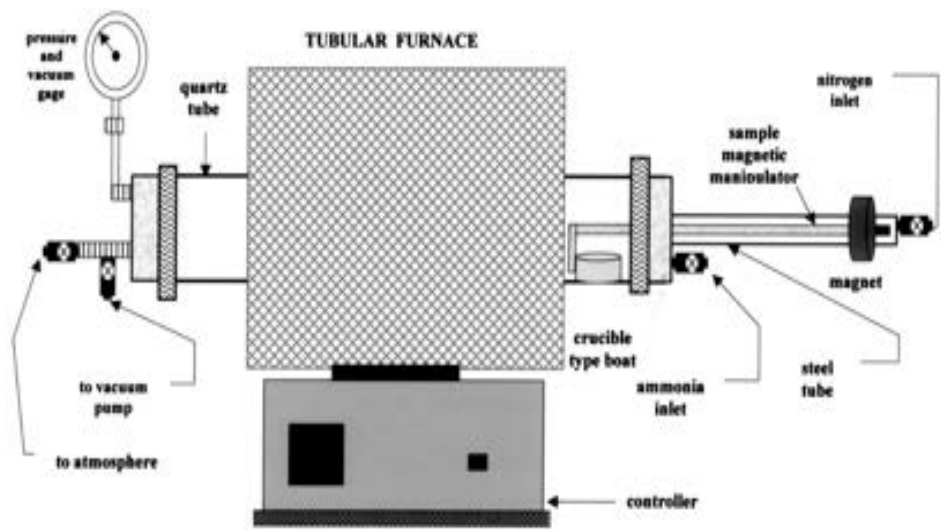

FIG. 3.6. A schematic representation of the tubular reactor.[68]

An alumina boat containing the dried precursor in powder form is placed at the entrance of the reactor, after which the reactor is closed. In order to remove the air from the reactor and to form an inert atmosphere, a flow of ultrahigh-purity nitrogen is supplied through the inlet valve at a rate of ~150 standard cubic centimeters per minute (sccm). The center of the reactor is heated until temperature reaches ~1000 °C while the entrance of the reactor where the alumina boat resides is maintained at ~160°C. Then, the inner part of the reactor is evacuated by a mechanical pump until reaching approximately 0.001 Torr. These conditions are maintained for ~1 hour in order to allow further drying of the ammonium hexafluoroindium-gallate and to reduce any contamination.

After 1 hour, the temperature at the center of the reactor is lowered to about 630°C and the nitrogen flow is closed. Simultaneously, ammonia is introduced in the reaction chamber at a flow rate of ~300 sccm. Then, using a magnetic manipulator, the boat with the ammonium hexafluoroindium-gallate is placed at the center of reactor (see FIG. 3.7).

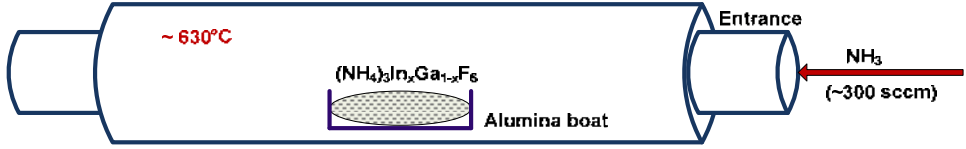

FIG. 3.7. A schematic of the quartz tube showing the flow of ammonia and the precursor at the center.

At this time the following reaction is carried out:

$$(NH_4)_3In_xGa_{1-x}F_6(s) + 4NH_3(aq) \rightarrow In_xGa_{1-x}N(s) + 6NH_4F(g) \qquad (3.11)$$

The reactor temperature is kept at 630°C for ~10 min and is then lowered to ~600°C and held at this temperature for another 25 min. The temperature is reduced again to ~520°C and kept for ~20 min. Finally, the boat is extracted to the entrance of the reactor using the magnetic manipulator. After the furnace is cooled, the sample is taken out for analysis.

# CHAPTER 4

# CHARACTERIZATION METHODS

## 4.1. X-RAY DIFFRACTION ANALYSIS

### 4.1.1. Basics of X-ray Diffraction

X-ray diffraction (XRD) is a versatile non-destructive structural characterization tool and is now a common technique for the study of crystal structures and atomic spacing. X-ray diffraction analysis can be used to determine identity, composition, crystal orientation, strain state, grain size, and epitaxial quality.

X-ray diffraction is based on constructive interference of monochromatic X-rays diffracted by a crystalline sample. These X-rays are generated by a cathode ray tube, filtered to produce monochromatic radiation, collimated to a concentrated beam, and directed toward the sample. The interaction of the incident x-rays with the sample produces constructive interference of the diffracted x-rays only at certain angles that satisfy Bragg's Law given by:

$$2d\sin(\theta_B) = n\lambda \qquad (4.1)$$

where $n$ is the order of reflection, $\lambda$ is the wavelength of the x-rays, $d$ is the interplanar spacing of the reflecting planes and $\theta_B$ is the Bragg angle. For any angle other than $\theta_B$ the scattered x-rays undergo destructive interference and hence no signal is perceived by the detector. Therefore, determination of $\theta_B$ reveals the interplanar spacing of the crystal phase, and in turn the lattice parameter. Fig. 4.1 gives an illustration of the atomic planes, the incident and diffracted x-rays.

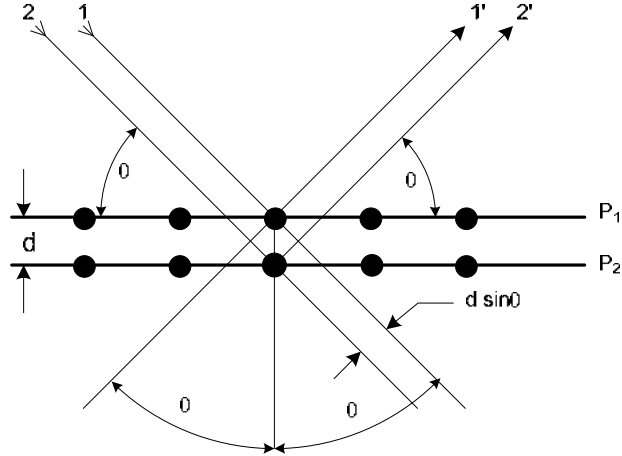

FIG. 4.1. The diffraction of incident X-rays by a pair of parallel atomic planes.

In Fig. 4.1. an X-ray beam is incident on a pair of parallel atomic planes $P_1$ and $P_2$, separated by an interplanar spacing $d$. The two parallel incident rays 1 and 2 make an angle $\theta$ with these planes. The diffracted rays 1' and 2' will interfere constructively when the difference in path length between 1-1' and 2-2' ($2d\sin\theta$) is an integral number of the wavelength $\lambda$. This relationship is expressed mathematically in Bragg's law given above in equation (4.1).

The intensity of the reflections is proportional to the square of the structure factor $F_{hkl}$, which is expressed in terms of the contents of a single unit cell as:

$$F_{hkl} = \sum_{j=1}^{N} f_j \exp[2\pi i(hx_j + ky_j + lz_j)] \qquad (4.2)$$

In equation (4.2) the position of the $j^{th}$ atom is given by the coordinates $(x_j, y_j, z_j)$, $f_j$ is the $j^{th}$ atom scattering factor, and $N$ is the total number of atoms in the unit cell. The atomic scattering factor $f_j$ is the Fourier transform of the charge density of electrons in the atom given by equation (4.3).

$$f(\vec{q}) = \int \rho(\vec{r}) e^{i\vec{q}\cdot\vec{r}} d^3\vec{r} \qquad (4.3)$$

where $\rho(r)$ is the electron charge density of the scattering atom, and $q$ is the momentum transfer. As can be seen from equations (4.2) and (4.3), the intensity of the diffracted x-rays depends on the kind of atoms and where in the unit cell they are located. Planes going through areas with high electron density will reflect strongly, planes with low electron density will give weak intensities. Furthermore, the areas under the peak are related to the amount of each phase present in the sample.

The diffracted X-rays are then detected, processed and counted. By scanning the sample through a range of 2θ angles, all possible diffraction directions of the lattice should be attained due to the random orientation of the powdered material. The powder diffraction method is thus ideally suited for characterization and identification of polycrystalline phases.

### 4.1.2. X-ray Diffraction Instrumentation

X-ray diffractometers consist of three basic elements: an X-ray tube, a sample holder, and an X-ray detector. X-rays are generated within a sealed tube that is under vacuum. An applied current heats a filament within the tube, the higher the current then the greater the number of electrons emitted from the filament. A high voltage, typically 15-60 kilovolts, is applied within the tube. This high voltage accelerates the electrons, which then hit a copper target. When electrons have sufficient energy to dislodge inner shell electrons of the target material, characteristic X-ray spectra are produced. These spectra consist of several components, the most common being $K_\alpha$ and $K_\beta$. $K_\alpha$ consists of $K_{\alpha 1}$ and $K_{\alpha 2}$, the former having a slightly shorter wavelength and twice the intensity of $K_{\alpha 2}$. In order to produce monochromatic X-rays needed for diffraction, monochromaters are used. $K_{\alpha 1}$ and $K_{\alpha 2}$ are sufficiently close in wavelength such that a weighted average of the two is used (~1.5418Å for Cu $K_\alpha$ radiation). These X-rays are collimated and directed onto the sample. As the sample and detector are rotated, the intensity of the reflected X-rays is recorded. When the geometry of the incident X-rays impinging the sample satisfies Bragg's law given in equation (4.1), constructive interference occurs and a peak in intensity occurs. A detector records and processes this X-ray signal and converts the signal to a count rate which is then output to a device such as a printer or computer monitor. An X-ray scan is done by changing the angle between the X-ray source, the sample, and the detector at a controlled rate between preset limits.

The geometry of an X-ray diffractometer is such that the sample rotates in the path of the collimated X-ray beam at an angle $\theta$ while the X-ray detector is mounted on an arm to collect the diffracted X-rays and rotates at an angle $2\theta$. The instrument used to maintain the angle and rotate the sample is termed a goniometer. For our powder samples, data is collected at angles $2\theta$ from ~20° to 80°.

A typical diffractometer has a fixed source of x-rays directed to the sample of interest and a detector on the other side of the sample to collect the diffracted beam. The sample and detector rotate to scan through a range of angles.

In this study, the X-Ray diffraction data were taken with a Rigaku D/Max-IIB x-ray diffractometer with Cu Kα radiation (wavelength ~ 1.5148 Å) in the Bragg – Brentano geometry (Fig. 4.2).

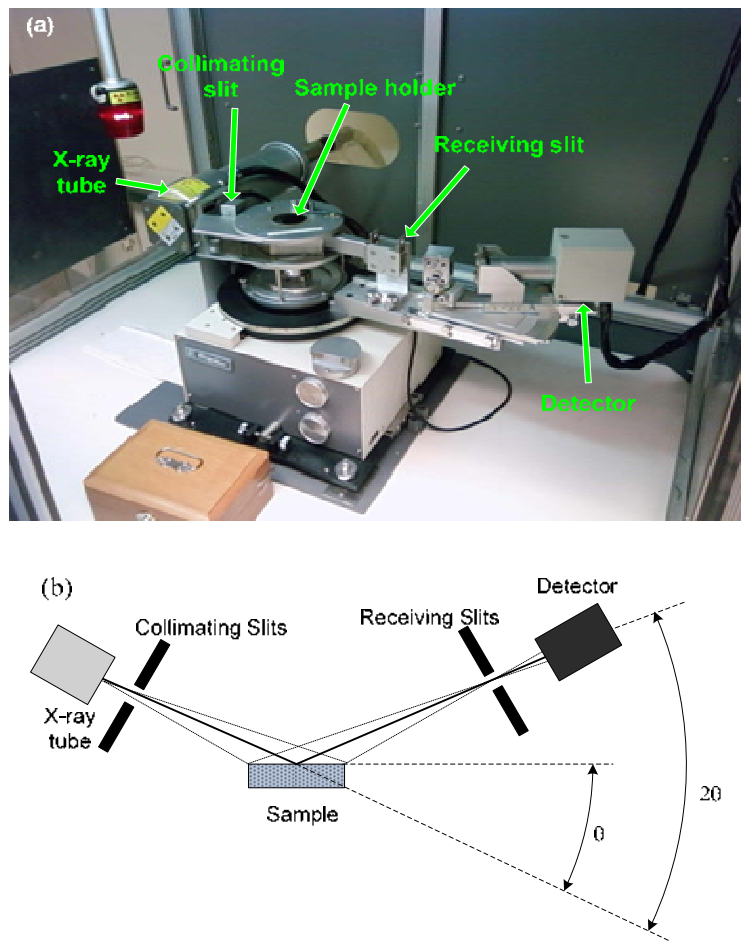

FIG. 4.2. (a) Photograph of the Rigaku D/Max-IIB X-ray diffractometer at ASU, (b) Schematic representation of the X-ray diffraction apparatus in the Bragg – Brentano geometry.

**4.1.3. Data Collection**

The intensity of diffracted X-rays is continuously recorded as the sample and detector rotate through their respective angles. A peak in intensity occurs when the material contains lattice planes with d-spacings appropriate to diffract X-rays at that value of $\theta$. Although each peak consists of two separate reflections ($K_{\alpha 1}$ and $K_{\alpha 2}$), at small values of $2\theta$ the peak locations overlap with $K_{\alpha 2}$ appearing as a hump on the side of $K_{\alpha 1}$. Greater separation of these peaks occurs at higher values of $\theta$. Typically, these combined peaks are treated as one. The position of the diffraction peak is typically measured as the center of the peak at 80% peak height. Results are commonly presented as peak positions at $2\theta$ and X-ray counts (intensity) in the form of a table or an x-y plot (Fig. 4.3.). The *d*-spacing of each peak is then obtained by solution of the Bragg equation for the appropriate value of $\lambda$.

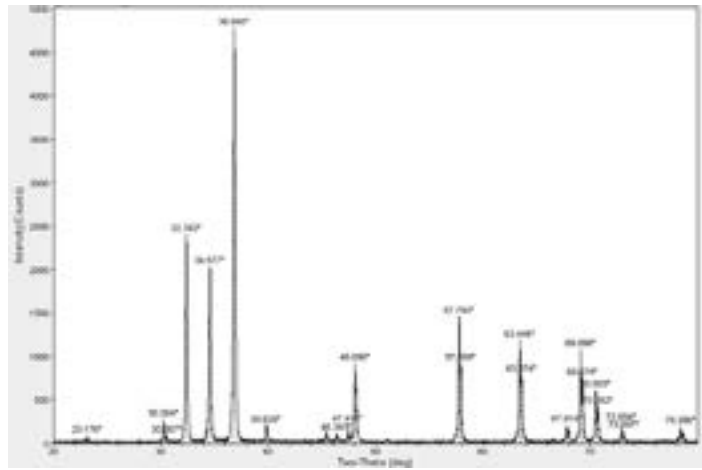

FIG. 4.3. An X-ray powder diffractogram. Peak positions occur at Bragg angles.

Knowing the values of interplanar spacings d, the lattice parameters of our materials can be found using equation (4.4) below.

$$\frac{1}{d_{hkl}} = \sqrt{\frac{4}{3}\frac{h^2 + k^2 + hk}{a^2} + \frac{l^2}{c^2}} \qquad (4.4)$$

In the above equation, *hkl* are the Miller indices of the atomic planes, *a* and *c* are the lattice parameters of the hexagonal structure. For the cubic structure, equation (4.5) is used to find the lattice parameter *a*.

$$d_{hkl} = \frac{a}{\sqrt{h^2 + k^2 + l^2}} \qquad (4.5)$$

In this study, $\theta$ - $2\theta$ scans were used routinely to determine the structure and composition of InGaN alloys. Composition *x* of the InGaN alloys can be estimated utilizing Vegard's Law given in equation (4.6).

$$c_{InGaN} = x c_{InN} + (1 - x) c_{GaN} \qquad (4.6)$$

All X-ray diffraction spectra are taken under the same conditions using 1° dispersion slits, 1° scatter slits and 0.15 mm receiving slit.

### 4.2. OPTICAL CHARACTERIZATION

Photon emission in the visible portion of the electromagnetic spectrum is an important property of semiconductors used for LEDs and LDs. These semiconductors have well defined energy bands and energy levels, and if enough energy is supplied, they readily emit light through the radiative recombination of electron – hole (e – p) pairs (see Fig. 4.4 (a)). The generated e – p pairs can move freely in the lattice and will eventually recombine generating a photon as shown in Fig. 4.4(b).

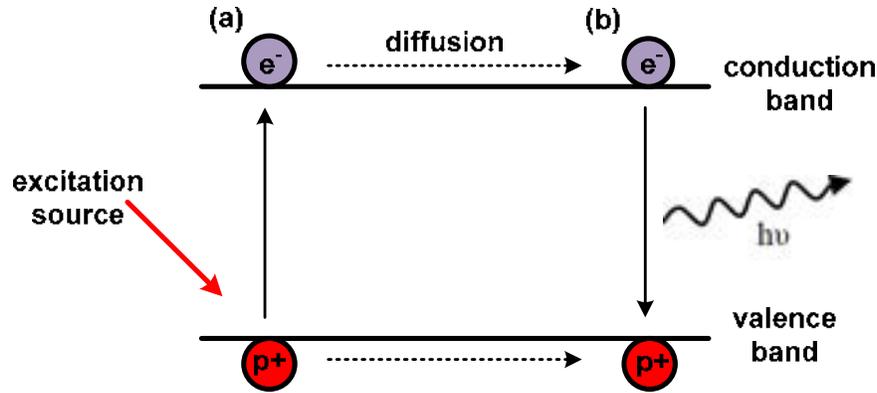

FIG. 4.4. Schematic representation of photon emission in semiconductors: (a) An external energy source incident on the material raises an electron to the conduction band leaving a hole in the valence band. (b) The generation of a photon after the recombination of carriers at a later time and position.

The excitation of the material for characterization of light emission can be done using one of the following techniques: photoluminescence, which makes use of a laser to excite carriers within the material; cathodoluminescence, which uses a high-energy electron beam to excite carriers; or electroluminescence, which uses an applied electric field to move carriers from doped regions and force recombination. In this study, the optical characterization of the materials is done using photoluminescence and cathodoluminescence techniques.

### 4.2.1. Photoluminescence

The emission of radiation induced by the optical excitation of a sample by using an external source of light is called *photoluminescence*. Photoluminescence (PL) is an effective and non – destructive method to evaluate the optical properties of wide band gap semiconductors. It can be used to characterize the energy band structure and radiative recombination centers. If a photon has energy greater than the bandgap energy of a semiconductor, then it can be absorbed and thereby raise an electron from the valence band up to the conduction band across the energy gap.

In this work, PL measurements are performed at room temperature using 325 nm He-Cd laser with an excitation power ranging from 3 mW to 10 mW, a spectrometer with diffraction grating of 1800 gratings/mm, and a Princeton Instruments UV model detector with a silicon charge-coupled device (CCD) camera.

### 4.2.2. Cathodoluminescence

The cathodoluminescence (CL) technique bombards a material with an incident electron beam of enough energy to generate e – p carriers. The use of electrons as the excitation source is an advantage of CL over other techniques, as the electrons can be finely focused to small dimensions, providing high spatial resolution. In many cases, this ability of CL is realized in a scanning electron microscope (SEM), which already possesses the electronics necessary to provide a nicely focused electron beam.

The electron beam energy used in CL measurements ranges from approximately 1 to 30 keV. These energies are large enough to induce ionization or create defects in certain materials. Nonetheless, the CL technique is in general considered non-destructive.

The CL system used in this dissertation contains a JEOL JSM 6300 scanning electron microscope with a $LaB_6$ filament, an Oxford MonoCL monochromator, and a Hamamatsu photomultiplier tube with GaAs detector for the collection of CL spectra. The monochromator grating is 1200 lines/mm blazed at 250nm in a Czerny-Turner configuration, with a dispersion of 2.7 nm per mm. SEM images were taken in conjunction with CL measurements. In our measurements we use a beam current of ~300 pA and an acceleration voltage of 5 kV.

**4.3. SCANNING ELECTRON MICROSCOPY**

The SEM uses a focused beam of high-energy electrons to generate a variety of signals at the surface of solid samples. The signals that derive from electron-sample interactions reveal information about the sample including external morphology (texture), chemical composition, and crystalline structure of materials making up the sample.

Accelerated electrons in an SEM carry significant amounts of kinetic energy, and this energy is dissipated as a variety of signals produced by electron – sample interactions when the incident electrons are decelerated in the solid sample. These signals include secondary electrons (that produce SEM images), backscattered electrons, luminescent photon (cathodoluminescence), characteristic X-rays, and Auger electrons (see Fig. 4.5.).

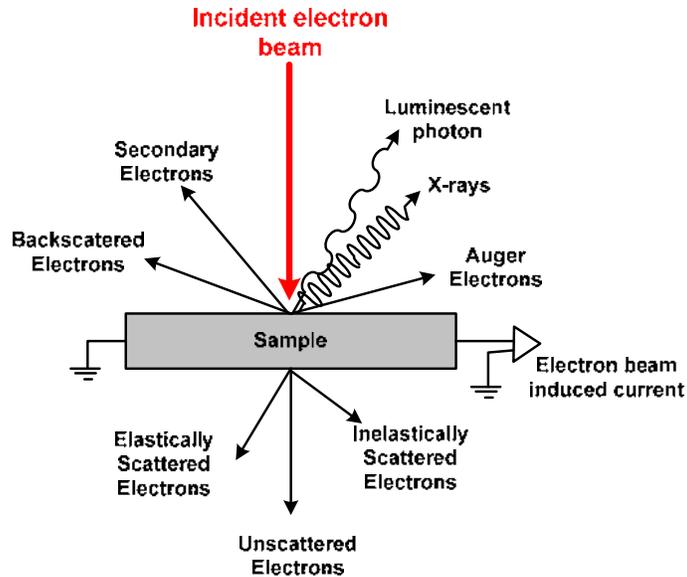

FIG. 4.5.   Schematic showing the possible physical processes in the SEM.

Image formation of surface structures using the SEM mainly depends on the production of secondary electrons, which happens by inelastic interactions of high-energy electrons with valence electrons as the incident beam is absorbed into the specimen within a few nanometers from the sample surface.  Backscattered electrons are high-energy electrons that are reflected or backscattered from the specimen by elastic scattering interactions with specimen atoms.  As a result of stronger backscattering from heavy elements, backscattered electrons provide chemical information about the sample. High-energy absorption leads to characteristic X-ray and Auger electron emission.  In the case of a thin sample, the incident beam can also be transmitted.  The SEM images in this dissertation are taken from an FEI XL30 environmental scanning electron microscope (ESEM) equipped with a field emission gun (FEG) and accelerating voltage of ~20kV.

# CHAPTER 5

# CHARACTERIZATION OF POWDERS GROWN FROM Ga-In MELT

## 5.1. INTRODUCTION

Bulk GaN crystal growth remains a technological challenge, but the desire to have large single crystals of GaN still continues. GaN growth on heterosubstrates often introduces unacceptably high defect densities. Due to the extreme melting conditions of GaN (~2500°C) and its relatively low decomposition temperature (~900 - 1100°C), it cannot be grown from its stoichiometric melt by the Czochralski method[70] commonly used for typical bulk semiconductor growth. This situation makes growth of bulk GaN a difficult task. To overcome such challenges and grow bulk GaN, several different techniques have been explored. GaN single crystals with low dislocation density have been prepared in Ga solutions under high $N_2$ pressure and high-temperature conditions[71, 72] and recently also by ammonothermal growth[73].

Other possible liquid phase GaN crystallization methods include melt and flux growth. Utsumi *et al*.[74] obtained GaN single crystals by direct cooling from the GaN melt under ultrahigh nitrogen pressures, above 60 kbar, and at temperatures close to 2200 °C. Unfortunately, these GaN crystals were small and of the quality not suitable for electronic applications[74]. Several attempts have been made to find a way for growing GaN from a liquid solution using various metal fluxes, mainly Na or Li, as solvents[75, 76]. The flux method has resulted in crystals several millimeters in size, but suffers strongly from the evaporation of the solvent due to its high vapor pressure and possible Na melt inclusions in the crystals. Moreover, this method still requires some pressure on the order of 5 MPa.

To avoid using a high pressure process it is necessary to choose an alternative nitrogen source, since Ga metal and nitrogen gas ($N_2$) do not react under atmospheric pressure. It has been shown that GaN could be grown under atmospheric pressure conditions using ammonia as the nitrogen source[77]. However, there are no reports of growing InGaN under these conditions. Here we investigate the possibility of growth of InGaN powders using the Ga-In melt under ammonia flow in atmospheric pressure.

## 5.2. RESULTS AND ANALYSIS

The procedure employed here for the synthesis of InGaN powders from the Ga-In melt is explained previously in section 3.4.1. This section will include the structural and optical characterization of the samples grown by this method. The attempted indium composition $x$ of these samples is given in Table 5.1.

Table 5.1. The attempted indium composition of three $In_xGa_{1-x}N$ samples.

| Alloy composition | Sample A | Sample B | Sample C |
|---|---|---|---|
| $x$ | 0 | 0.11 | 0.19 |

The growth of GaN (sample A) was attempted at three different temperatures 900°C, 1000°C and 1100°C. A reaction was observed to occur only at 1100°C (1 hr) when a gray color powder was formed. A characteristic feature of this reaction is the deposition of GaN powders on the inside and outside walls of the boat. At lower temperatures (T < 1100°C) no apparent reaction occurs. This is due to the formation of a thin crust of GaN over the melt which blocks the diffusion of nitrogen atoms into the molten Ga and hence no further reaction occurs.

Fig. 5.1 shows the XRD pattern of the powder in θ-2θ scan mode. The powder exhibits sharp peaks located at ~32.37°, 34.55° and 36.81° corresponding to the (10$\bar{1}$0), the (0002), and the (10$\bar{1}$1) planes, respectively, of wurtzite GaN. The lattice parameters are calculated to be c = 0.5192 nm and a = 0.3193 nm (c/a = 1.626), which are in excellent agreement with the reported values for hexagonal GaN[78, 79].

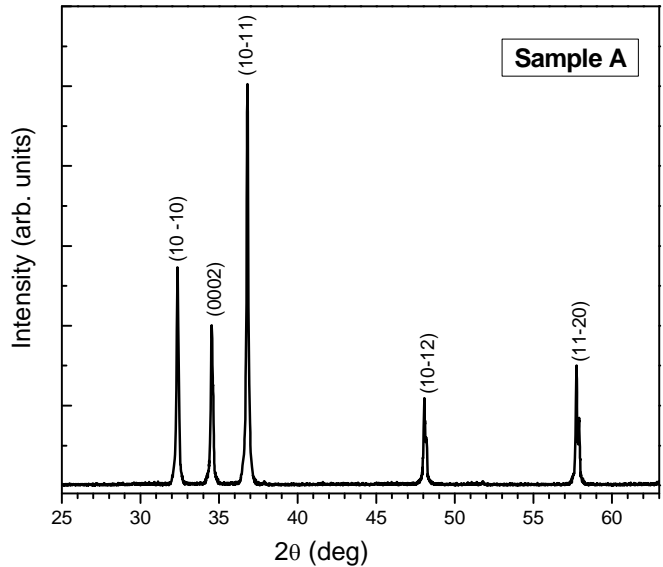

FIG. 5.1. XRD of sample A grown by the reaction of Ga metal with NH$_3$ at 1100°C under atmospheric pressure.

No peaks associated with metallic Ga appear in the XRD pattern shown in Fig. 5.1. This indicates that there are no Ga inclusions in the powder. The elemental composition of the GaN powder was determined by energy dispersive x-rays (EDX). Fig. 5.2 is an EDX spectrum of sample A and shows only peaks corresponding to Ga and N. The fact that there are no other transitions indicates the absence of impurities such as oxygen within the detection limits of the instrument.

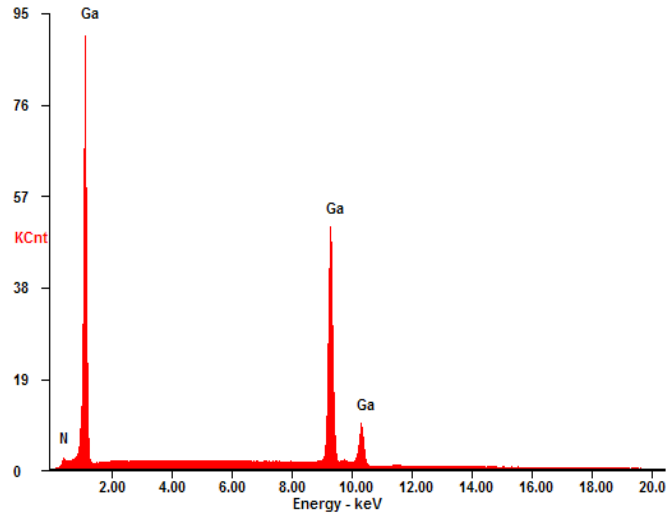

FIG. 5.2.  An EDX spectrum of GaN powder grown by the reaction of Ga melt with $NH_3$.

Secondary electron images of the obtained GaN powder are shown in Fig. 5.3. The powder is composed of small crystals ~ 5μm in size. These crystals appear in the form of platelets with clear hexagonal symmetry as shown in Fig 5.3(a).  In addition to the platelets, we also observe some small crystals.  Fig. 5.3 (b) shows that these crystals vary in size and have clear facets.

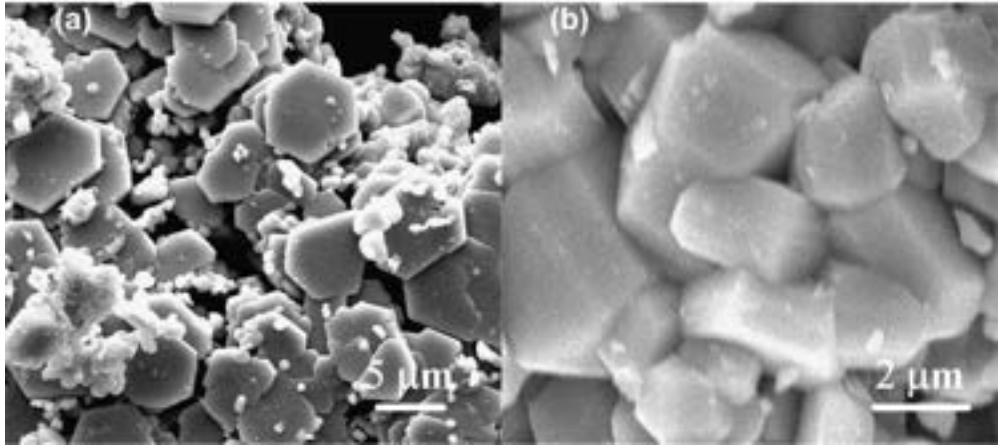

FIG. 5.3. Secondary electron images of the GaN powders. (a) Image shows platelet shape crystals with hexagonal symmetry. (b) A magnified image of the small crystal clusters on top of the platelets.

The optical properties of the synthesized GaN powder were investigated using PL at room temperature. The PL spectrum in Fig. 5.4 shows a strong near band-edge (NBE) emission at ~3.38 eV without any longer wavelength emission such as the yellow emission (~2.2 eV) often observed in undoped GaN. This excellent luminescence property of the powder indicates the growth of high quality GaN.

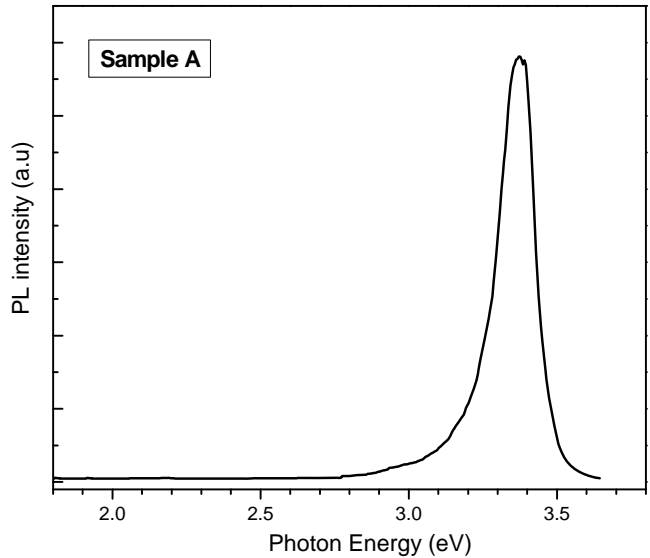

FIG. 5.4. Room temperature PL spectrum of GaN powder, showing near band-edge emission and no longer wavelength luminescence.

The growth of samples B and C was attempted at temperatures of ~800 °C, 900 °C, 1000 °C, and 1100 °C. As with the growth of GaN, only a dark gray crust was formed for temperatures lower than 1100°C, and most of the Ga-In alloy remained unreacted in the liquid form even after ~5 hr of reaction time. When the temperature of zone 2 (Fig. 3.7) was raised to ~1100°C, the reaction was completed in ~15 min when a dark gray powder was formed.

Fig. 5.5 shows the θ-2θ scans of samples B and C respectively. Both samples show similar XRD pattern. Sample B exhibits peaks located at ~32.36°, 34.54°, and 36.84° corresponding to the ($10\bar{1}0$), the (0002), and the ($10\bar{1}1$) planes of the wurtzite GaN, respectively, as seen in Fig 5.5(a). Similar peaks located at ~32.34°, 34.55°, and 36.83° are observed for sample C (see Fig. 5.5(b)). In addition, peaks associated with metallic In are also observed for both samples, while no peaks associated with metallic Ga, InN, or InGaN appear in either XRD pattern.

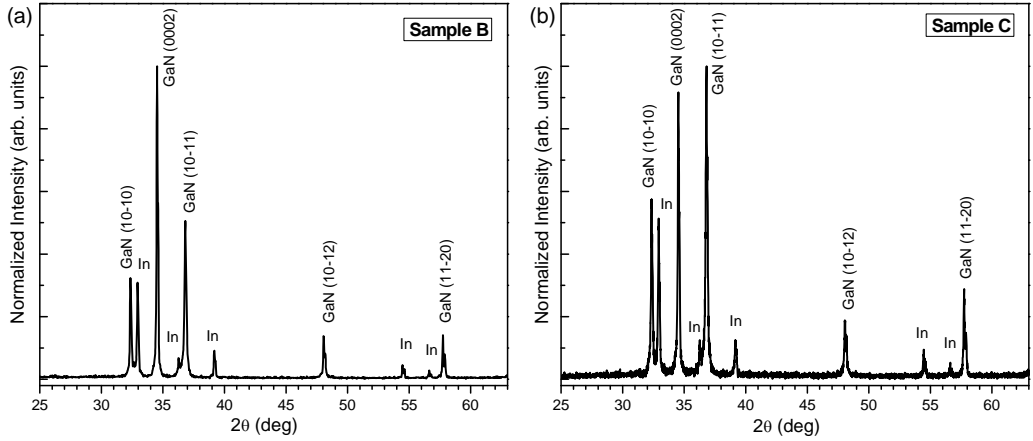

FIG. 5.5. XRD patterns of samples B and C, showing the hexagonal structure of GaN with metallic In inclusions.

The calculated values of lattice parameters of the samples grown by the reaction of the Ga-In melt with ammonia are given in Table 5.2.

Table 5.2. Lattice parameters of powders grown from Ga-In melt.

| Sample | a (nm) | c (nm) | c/a |
|--------|--------|--------|-------|
| A | 0.3193 | 0.5192 | 1.626 |
| B | 0.3194 | 0.5193 | 1.626 |
| C | 0.3196 | 0.5192 | 1.625 |

All three samples show the hexagonal structure of GaN with very small differences in the lattice parameters indicating that the structural properties of the GaN present on the samples grown by this method is not affected by the In concentration in the Ga-In melt.

Using the Scherrer equation ($d = K\lambda/\beta\cos\theta$),[80] the average size of GaN crystallites for sample B and C is calculated to be 66 nm and 62 nm respectively, compare to 65 nm found for pure GaN powder grown by the same method. This suggests that the crystallite size of GaN does not depend on the In content used for its growth.

EDX spectra for samples B and C are shown in Fig. 5.6(a) and (b) respectively. Only peaks of Ga, In, and N are observed. Both samples appear to be free of impurities such as C and O within the detection limits of the instrument.

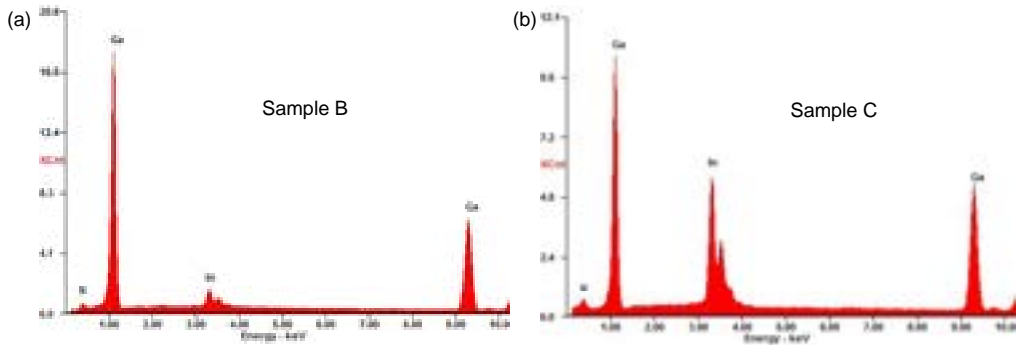

FIG. 5.6.  EDX spectra of samples B (a) and C (b), showing only peaks of Ga, In, and N.

The EDX spectra also show that the intensity ratio of In/Ga is higher in sample C than that in sample B, indicating that the In metal concentration in sample C is higher, as we would expect.

Fig. 5.7 shows secondary electron images of samples B (Fig 5.7(a)) and C (Fig 5.7(b)). Both samples form platelets with hexagonal symmetry ~15 μm in diameter. A property of these platelets, not observed in sample A (pure GaN), is the formation of tiny pits less than 1μm in diameter. In addition to these small pits, we also observe some bigger pits ~ 2 μm in diameter, with sample C exhibiting more of them per unit area. Smaller crystallites ~1μm in size with clear hexagonal symmetry (Fig. 5.7(c)) are also observed in both samples.

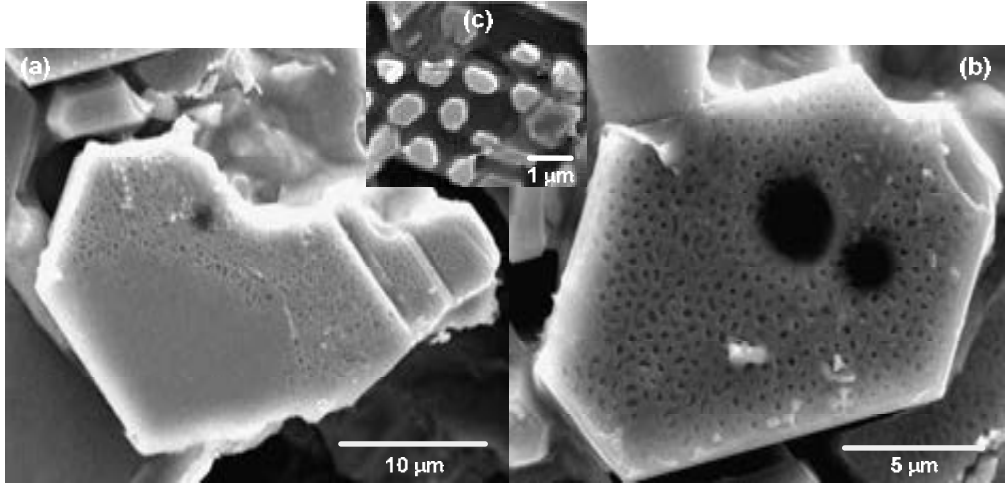

FIG. 5.7.  Secondary electron images of sample B (a) and sample C (b), showing pitted platelets with hexagonal symmetry. (c) Image of smaller crystals observed in both samples.

Room temperature PL spectra of samples B and C are shown in Fig. 5.8. Both spectra exhibit a sharp peak associated with NBE luminescence of hexagonal GaN. This peak is located at ~3.38 eV for sample B and at ~3.36 eV for sample C. A broader luminescence peak is also observed in both spectra, but with higher intensity for sample C. These peaks are centered at ~2.52 eV and 2.41 eV for samples B and C respectively. Similar broad green emissions have been reported for undoped[81] and slightly n-type[82] GaN. It has been suggested that this band is related to isolated native defects such as Ga vacancies ($V_{Ga}$) or its complexes with oxygen.[81, 83] Even though the EDX spectra for samples B and C show no presence of impurities, it is possible that O may be a contaminant in small amount (less than 1%).

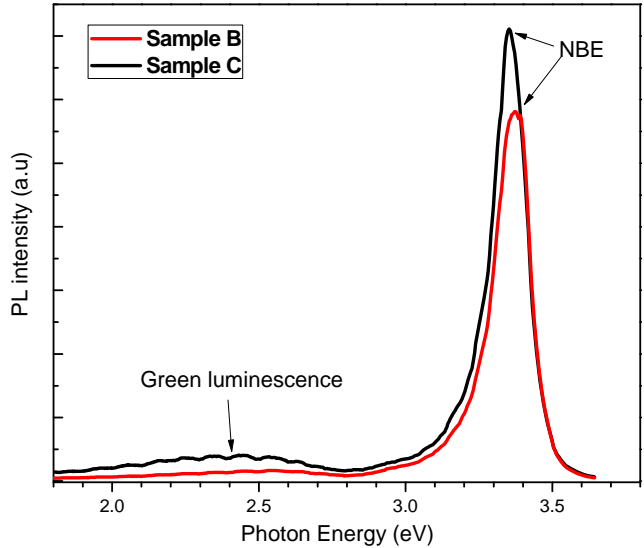

FIG. 5.8. Room Temperature PL spectra of samples B and C, showing near band-edge luminescence of hexagonal GaN and a broad green luminescence.

These PL results show that the green band emission is suppressed for sample A (pure GaN) where In is absent during growth. This is in agreement with reported calculations that Ga vacancies are less likely to form under Ga-rich conditions.[84] The ratio of Ga/N decreases from sample A to C as the In concentration in the melt is increased, while the flow rate of $NH_3$ remains the same. This also explains the increase in intensity of the green luminescence as the In concentration is increased (see Fig. 5.9). The formation of $V_{Ga}$ becomes more favorable moving away from Ga-rich conditions, spurring an increase in intensity of this band.

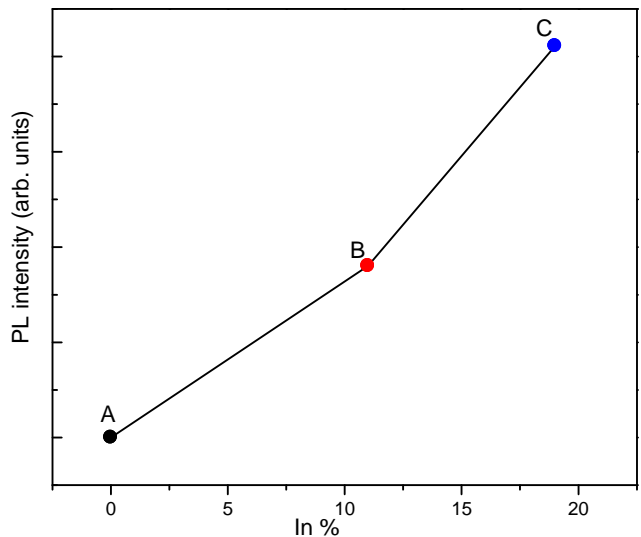

FIG. 5.9. A graph of PL intensity of the green emission versus In composition in Ga-In melt.

## 5.3. KINETIC AND THERMODYNAMIC CONSIDERATIONS OF THE GROWTH

Indium nitride is the most difficult of the III-nitrides to grow due to the high equilibrium vapor pressure of nitrogen and the low decomposition temperature of solid InN (~500 °C).[85, 86] The curves in Fig. 5.10 represent the equilibrium nitrogen pressures versus temperature for AlN, GaN, and InN. They indicate that at $N_2$ pressure of ~1 bar, GaN is thermodynamically stable up to temperatures approaching ~ 1200 K, whereas InN loses its thermodynamic stability already at temperatures as low as ~ 580 K. It can also be seen that the application of higher $N_2$ pressures permits significant extension of the stability range of both GaN and InN, but with a much smaller effect on InN. These data are in agreement with the experimental results presented in the previous section. According to Fig. 5.10, the successful growth of InN for nitrogen pressures below

~ 1bar (used in the synthesis of the powder samples in this study), would require a growth temperature of ~625 K, while GaN cannot be grown under these conditions. Increasing the temperature to ~1000 K makes GaN growth stable, but destabilizes InN.

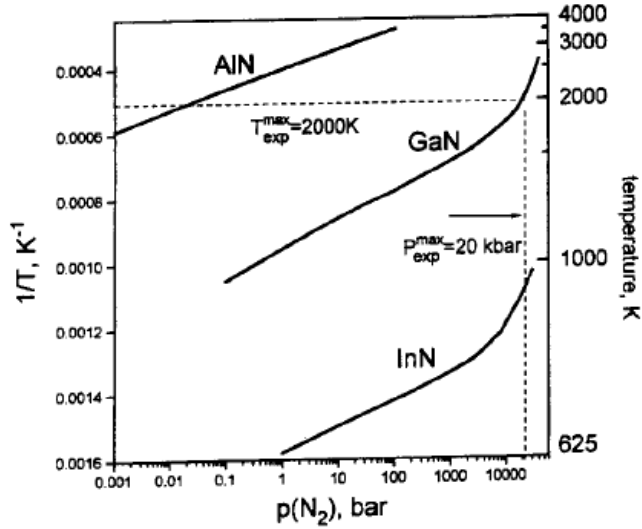

FIG. 5.10. Equilibrium $N_2$ pressure over III-V nitrides.[85]

At high temperatures InN decomposes[87] according to the reaction:

$$InN(s) \rightarrow In(l) + \frac{1}{2} N_2(g) \tag{5.1}$$

This explains why InGaN/InN was not present in our samples and only In metal could be found. It is possible that InN formed in the crust of the Ga-In melt at lower temperatures, but decomposed as the temperature was increased. Experimental observations show that a film is present on the surface of the metallic alloy at lower temperatures, but its identification by x-ray techniques was not possible. Evidence for the presence of InN on the crust could be the formation of some gas bubbles as the temperature is increased. These gas bubbles are not observed when only Ga metal is used. This result leads to the suggestion that the bubbles could be $N_2$ escaping from the decomposition of InN. Another possibility is that the nitride formed on the metallic surface greatly slows down the diffusion of nitrogen through this film at the temperatures investigated. At any rate, it appears that the kinetics of formation of InGaN from Ga-In and $NH_3$ are extremely unfavorable at low temperatures where the InN phase is stable.

In order to further understand the role of kinetics in the formation of GaN and InN from the Ga-In alloy, an analysis of growth process is done from this point of view. First, the dissociation of $NH_3$, which follows according to equation (5.2) below, is an important issue to consider during the growth process.

$$NH_3 \rightarrow \frac{1}{2}N_2 + \frac{3}{2}H_2 \tag{5.2}$$

It is found that $NH_3$ dissociation is dependent on temperature and is increased by one to two orders of magnitude on a growing III-N surface.[87] The dependence of ammonia dissociation on temperature during GaN growth is displayed in Fig. 5.11, which shows that there is very little $NH_3$ dissociation at low substrate temperatures. The dissociation rate is strongly enhanced when the substrate temperature is increased. $NH_3$ dissociation reaches a maximum at higher temperatures and then begins to decline, due to the onset of gallium desorption.

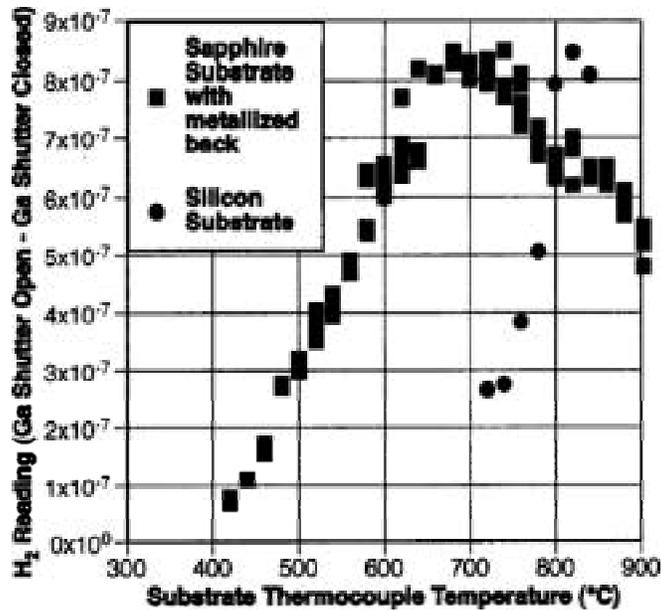

FIG. 5.11. $H_2$ readings in Torr ($NH_3$ dissociation) during GaN growth versus substrate thermocouple reading, showing the thermocouple offset between silicon and backside-metallized sapphire substrates.[88]

The dissociation of ammonia leads to the production of $N_2$ and $H_2$ according to equation (5.2). The role of hydrogen on III-nitride growth has been studied,[89, 90] since hydrogen is present in high concentrations in commonly used growth techniques for nitride semiconductors such as MOVPE, HVPE, and MBE using an $NH_3$ source. Hydrogen has been observed to have important effects on the growth of GaN, such as an increase of the growth rate by a factor of two[89] and also improvement of the quality of GaN.[90] On the other hand, $H_2$ is believed to be detrimental to the incorporation of indium in InGaN.[91] InN bonds are less stable and dissociate more readily than GaN bonds, due to their lower energy (~7.72 eV) compare to that of GaN (~8.92 eV). Therefore, $H_2$ molecules might bond with a surface N atom in the InN bond, producing a $NH_2$ free radical and leaving metallic indium on the surface.

The adsorption of diatomic molecules, such as molecular nitrogen $N_2$, on group III metal surfaces leads to dissociation of the molecule in the adsorption stage and attachment of single N atoms at the surface.[92] The dissociation of $N_2$ is characterized by relatively large energy barriers ~3.4 eV for Ga and ~3.6 eV for In. The process of dissociation is also dependent on the orientation of the nitrogen molecule; for parallel orientation of $N_2$ molecule the energy barrier increases to ~4.8 eV for Ga and ~5.8 eV for In. Using ideal gas approximation for $N_2$, and assuming only perpendicular direction of $N_2$, its dissociation rate $R$ is given by:[92]

$$R = \int v f(v) P(E) dv \quad (5.3)$$

where $v$ is the velocity of the gas molecules, $f(v)$ is the Maxwell-Boltzman distribution function of the velocities of $N_2$ gas, and $P(E)$ is the probability of penetration of a molecule with energy $E$ through the energy barrier $E_b$. Since only perpendicular direction of $N_2$ is assumed, then the total energy of the molecule is the sum of translational and vibrational energies ($E_t$ and $E_v$). The probability of $N_2$ to overcome the energy barrier is given by:

$$P(E) = \begin{Bmatrix} 1 \text{ for } E > E_b \\ 0 \text{ for } E < E_b \end{Bmatrix} \quad (5.4)$$

where $E$ is $E_t$, $E_t+E_v$, or $E_v$.

The Maxwell-Boltzmann function for the distribution of velocities (only one direction) is given by:

$$f(v) = \sqrt{\frac{m}{2\pi kT}} \, e^{-\frac{mv^2}{2kT}} \quad (5.5)$$

After substitution of (5.4) and (5.5) into (5.3), using the ideal gas equation $pV = NkT$ and the approximation $E_b \gg kT$, the dissociation rate $R$ takes the form:

$$R = \frac{p_{N_2}}{\sqrt{2\pi mkT}} (1 + \frac{E_b}{kT}) e^{-\frac{E_b}{kT}} \tag{5.6}$$

where $p_{N2}$=16,900 *Pa* is the partial pressure of $N_2$, which is found from multiplying total pressure *p* (~101,325 *Pa*) with the molar composition of $N_2$ from the dissociation of ammonia. The dissociation rate of nitrogen on Ga and In metal surfaces is plotted in Fig. 5.12 using $E_b$ values of 3.4 eV and 3.6 eV for Ga and In respectivily. Fig. 5.12 shows that, at a given temperature, the rate of dissociation of molecular nitrogen on gallium surface is higher by an order of magnitude than that on In surface. For a Ga-In alloy we would expect the $N_2$ dissociation rate to be lower than in the case of gallium, because the energy barrier increases by the addition of indium in the melt.

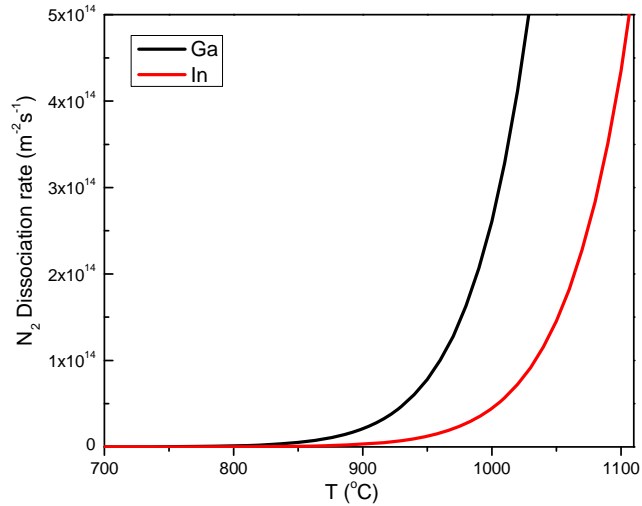

FIG. 5.12. The dissociation rate of $N_2$ on Ga and In melt surface.

This increase in the kinetic barrier of $N_2$ dissociation on liquid Ga-In surface may be one of the reasons the growth of InGaN was proven difficult for our experimental conditions. Even though the dissociation rate of $N_2$ on the In surface increases with temperature, the low thermal stability of InN prevents the effective synthesis of InN at higher temperatures.

When $N_2$ dissociates on molten Ga or Ga-In surfaces, the atomic nitrogen begins to diffuse into the melt, as the nitrogen concentration at the surface increases. Fick's second law of diffusion for non-steady state can be applied to find the concentration of nitrogen in the melt $C(x,t)$ at a given depth $x$ from the surface and at a given time $t$:

$$\frac{\partial C(x,t)}{\partial t} = D \frac{\partial^2 C(x,t)}{\partial x^2} \tag{5.7}$$

where $D$ is the diffusion coefficient of nitrogen in the melt. Assuming that $D$ is independent of position and using boundary conditions $C(0,t) = C_s$ and $C(x,0) = C_0$, the solution to the above equation is:[93]

$$C_x = C_s - (C_s - C_0)\mathrm{erf}\left(\frac{x}{2\sqrt{Dt}}\right) \tag{5.8}$$

where $C_x$ is the concentration of nitrogen at distance $x$ from the surface at time $t$, and *erf* is the error function given by:

$$\mathrm{erf}(z) = \frac{2}{\sqrt{\pi}} \int_0^z e^{-u^2} du \tag{5.9}$$

A plot of concentration of nitrogen in the melt versus depth $x$ for two different times is shown in Fig. 5.13. While nitrogen concentration remains at the value of $C_s$ on the surface of the melt, it increases for a certain depth as time increases.

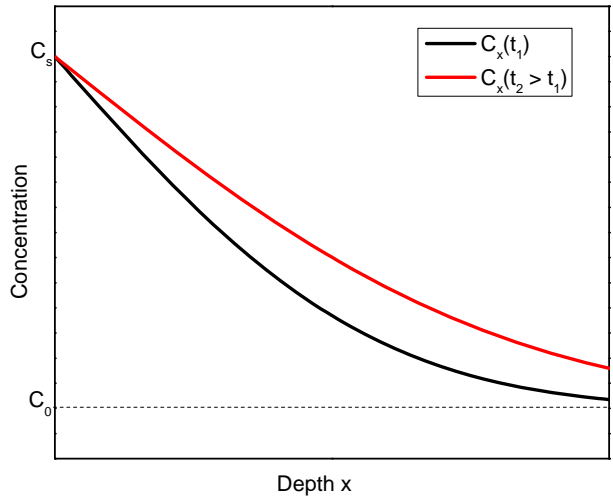

FIG. 5.13. Concentration profiles of nitrogen at various times versus distance below the surface of the melt. The melt had an initial concentration of nitrogen $C_0$ before diffusion started.

Formation of InGaN (crystallization) will occur when the nitrogen concentration in the melt reaches the saturation level, which happens earlier at the surface. As a result, a crust is formed as observed during growth. In Fig. 5.14 concentration of nitrogen is plotted as a function of depth for $C_0 = 0$ and $C_0 \neq 0$ at the same time $t$ after diffusion starts. The profiles in Fig. 5.14 show that the concentration of nitrogen at a certain depth $x$ in the melt is higher for $C_0 \neq 0$, which means that saturation is reached faster than in the case of $C_0 = 0$. This provides an insight into the fact that the growth of GaN proceeds very slowly when the solution of Ga-In-$N_2$ is not made prior to the synthesis, in which case only a crust is formed even after ~12 hr of time growth. By initially diluting the melt with nitrogen the growth process is accelerated and the complete conversion into GaN happens in ~1 hr. The growth is even faster (~15 min) when In is present in the melt solution.

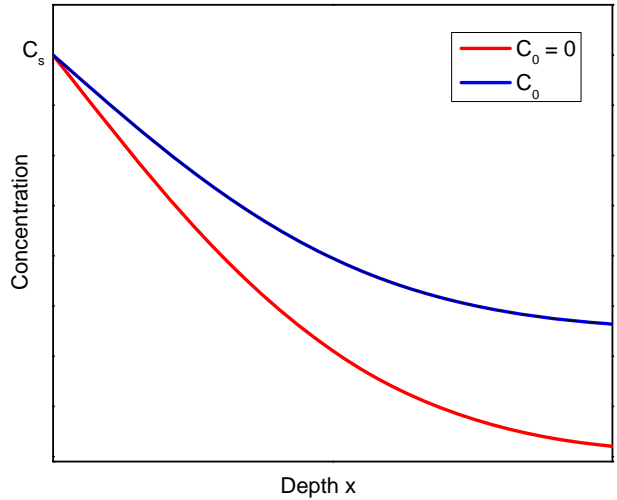

FIG. 5.14. Concentration profiles of nitrogen for zero and non-zero initial concentration $C_0$ (same time after diffusion starts).

It is also important to consider the fact that the diffusion coefficient $D$ increases with temperature ($\propto exp(-E_D/RT)$), which also causes an increase of the concentration of nitrogen in the melt. This would also explain why the GaN formation is completed faster at higher temperatures.

Thermodynamics considerations can also help in the understanding of the growth process. The InGaN alloy forms from the reaction of Ga and In atoms with the dissolved nitrogen atoms in the melt according to the following equations:

$$Ga(l) + N(g) \rightarrow GaN(s) \qquad (5.10)$$

$$In(l) + N(g) \rightarrow InN(s) \qquad (5.11)$$

where GaN (s) and InN (s) are the binary compounds in the InGaN alloy. The equilibrium condition for GaN requires that:

$$\mu_{Ga}^0 + \mu_N^0 = \mu_{GaN} \qquad (5.12)$$

When In is added to the Ga melt, the partial chemical potential of Ga in the alloy is found by the derivative of Gibbs free energy according to the following:

$$\mu_{Ga} = \left(\frac{\partial G}{\partial n_i}\right)_{T,p,n_{In}} \quad (5.13)$$

By using the Gibbs free energy for the regular solution model of a binary alloy derived in chapter 2 the partial chemical potential for Ga is found to be:

$$\mu_{Ga} = \mu_{Ga}^0 + RT\ln(1-x) + \Omega x^2 \quad (5.14)$$

where $x$ is the composition of In in the Ga-In alloy. Fig. 5.15 shows a plot of the change in the chemical potential of Ga in the alloy as a function of the In composition for different temperatures, where a value of 1060 *cal/mol* is used for $\Omega$ of Ga-In alloy.[94]

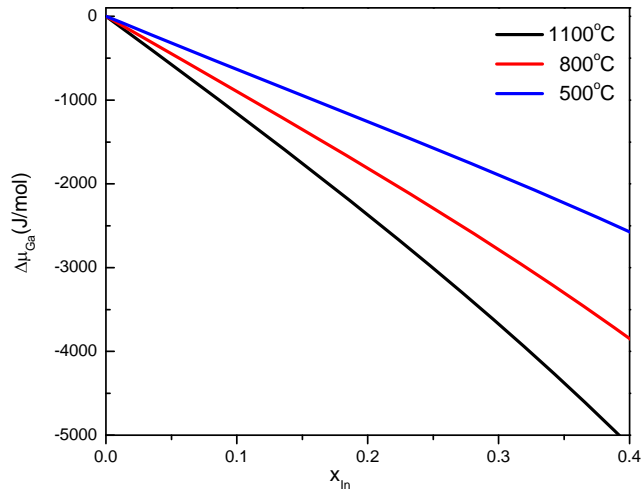

FIG. 5.15. The change in chemical potential of Ga as a function of In composition in the Ga-In alloy for various temperatures.

The chemical potential of Ga decreases by the addition of indium in the melt. Also, this decrease is more pronounced at higher temperatures. Since the equilibrium solubility of InN in GaN is very small,[59] the chemical potential of GaN produced by the Ga-In alloy doesn't change significantly compare to the GaN produced by the pure Ga melt. Equation (5.12) can be rewritten:

$$\mu_{Ga}^0 + \mu_N^0 = \mu_{Ga} + \mu_N \tag{5.15}$$

According to equation (5.15) the chemical potential of nitrogen dissolved in the Ga-In alloy must be higher than that in the pure Ga melt. As a result the nitrogen concentration is higher, which causes saturation, and in turn crystallization, to happen much faster than in the case of pure Ga. This gives a thermodynamic explanation of the faster GaN growth in the case of the Ga-In melt. It also explains why only a crust is formed at lower temperatures where the nitrogen concentration is not high enough to complete the crystallization. The increase in nitrogen concentration also causes a reduction on the Ga/N ratio, which is believed to be a factor in the formation of $V_{Ga}$[84] as mentioned in the previous section. This explains the increase in intensity of the green emission of GaN when indium composition in the metallic alloy increases as observed in the PL spectra.

## 5.4. CONCLUSIONS

The growth of InGaN powders was attempted by using the reaction of molten Ga-In alloy with ammonia in a horizontal tube reactor. X-ray diffraction measurements indicate the hexagonal structure of GaN without any InN or InGaN alloy present. Characterization of the samples by PL shows the NBE emission of GaN, with a weak broader deep level emission, whose intensity increases with the indium content in the molten alloy. The growth of InN in our samples, either as a separate phase or as an alloy with GaN, was shown to be unsuccessful. This can be attributed to the low nitrogen pressures used compare to the equilibrium vapor pressure of nitrogen and also to the low decomposition temperature of solid InN. Kinetic and thermodynamic considerations can be used to explain the growth process.

# CHAPTER 6

# CUBIC GaN CRYSTALLITES WITH SHARP LUMINESCENCE LINES

## 6.1. INTRODUCTION

GaN and its alloys form direct band gap semiconductors, capable of emitting light across the visible spectrum and into the ultraviolet range.[1] The development of GaN based optoelectronic devices has become an important field at the forefront of semiconductor research. High quality GaN films have been grown epitaxially with the hexagonal wurtzite structure by metal organic chemical vapor deposition (MOCVD) on sapphire substrates, and are used commercially in the fabrication of light emitting devices.[95–98] GaN films have also been grown with the less energetically favorable cubic (zincblende) structure.[99–101] In certain cases both structures may coexist because of their small difference in energy of formation.[102–104]

Cubic GaN offers many potential advantages over hexagonal GaN. The built-in electric fields arising from the spontaneous and piezoelectric polarizations are considerable in hexagonal GaN heterostructures[105, 106] and deemed undesirable for light-emitting devices. The absence of these fields in cubic (100) GaN, due to a center of symmetry in the lattice, is highly desirable for device application. Also, phonon scattering is lower in the cubic phase due to the higher crystallographic symmetry.[107] As a result the mobility of electrons and holes in the cubic phase is higher than in the hexagonal one.[108] Cubic GaN has a smaller energy band gap of ~ 3.3 eV at 5 K[108–110] compare to ~3.4 eV for the hexagonal phase. This reduces the indium composition required for green light emission, which is still a challenge in wurtzite III-nitride based optoelectronic devices.

In this work, we have investigated the use of a chemical vapor deposition (CVD) reactor to grow GaN thin films on amorphous silicon dioxide substrates. Characterization by X-ray diffraction (XRD) and cathodoluminescence (CL) shows that the film is primarily hexagonal in structure. Small inclusions of cubic phase GaN have been detected, exhibiting remarkably sharp peaks and strong luminescence.

## 6.2. EXPERIMENTAL PROCEDURE

A schematic diagram of the CVD reactor used in this study is shown in Fig. 6.1. It consists of a horizontal 2.5″ diameter quartz tube, heated by a resistance furnace, with two inner quartz tubes and three separate thermal zones.

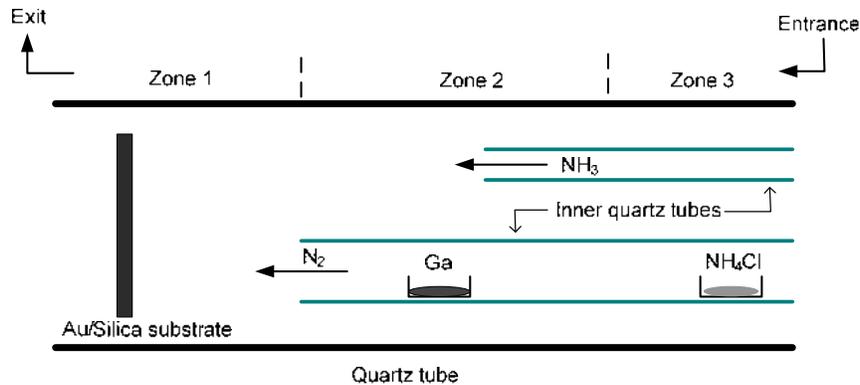

FIG. 6.1. Schematic diagram of horizontal tube reactor for growth of GaN by CVD.

A boat holding a fused-silica substrate in a vertical position is located in zone 1. An inner quartz tube, with a diameter of ¾″, holds a boat filled with liquid gallium located in zone 2, and a second boat containing ammonium chloride salt (NH₄Cl) in zone 3. Ammonia gas (NH₃) is fed into zone 2 by the second quartz tube. The silica substrate was first covered with a gold layer with a thickness of ~10nm, by physical vapor deposition, and it was subsequently annealed at 800°C. During annealing, the gold film coalesced into droplets with a diameter of ~ 1 μm, as previously reported for growth of GaN films.[111] The droplets are believed to act as nucleation sites for growth of GaN by a vapor-liquid-solid mechanism.

Prior to deposition, the boats containing the substrate and the precursors were placed into position using magnetic manipulators, and the entire system was evacuated. The reactor was then flushed with nitrogen gas. The temperatures in zone 1 and zone 2 were raised to ~900°C and 800°C respectively. At this time the flow of ammonia gas was initiated at ~500 standard cubic centimeters per minute (sccm). Evaporation and decomposition of the ammonium chloride salt began when zone 3 reached ~300°C according to the following equation:

$$NH_4Cl(g) \rightarrow NH_3(g) + HCl(g) \tag{6.1}$$

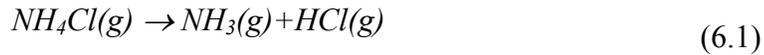

A flow of nitrogen gas of ~40 sccm transported the ammonia and hydrochloric acid vapors down the tube. Hydrochloric acid reacts with the gallium melt to produce gallium chloride (GaCl₃). The latter is carried out toward the substrate.

$$Ga(l) + HCl(g) \rightarrow GaCl_3(g) \tag{6.2}$$

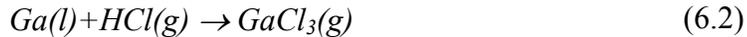

The growth of GaN occurs at the substrate from the reaction of GaCl₃ with ammonia at a higher temperature until NH₄Cl is consumed.

$$GaCl_3(g) + NH_3(g) \rightarrow GaN(s) \tag{6.3}$$

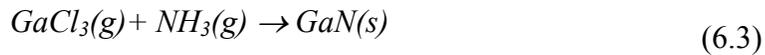

The gas flows were stopped, and the substrate with the newly deposited film was removed from the hot regions of the furnace and allowed to cool down. The growth process typically lasts about one hour. Two GaN films were grown at ~900°C (sample A) and ~800°C (sample B).

## 6.3. RESULTS

Secondary electron (SE) images, such as the plan-view images in Fig. 6.2(a) and (c), show a polycrystalline film with a primarily hexagonal columnar structure in both samples, but the morphology differs. The columnar structure in sample A appears to be randomly oriented with respect to the substrate with a large variation in column height and width. Cross-section SEM images such as Fig. 6.2(b) show that the larger columns may protrude as much as 5-6 μm above the substrate. Smaller crystallites appear in a layer densely packed near the substrate, lack apparent symmetry, and vary in size discernable down to the submicron range. The hexagonal structure of sample B (Fig. 6.2(c)) has a preferred [0001] orientation with crystallites of ~1 – 2 μm in diameter. The SEM image of the cross-section in Fig. 6.2(d) shows a coalesced film with sharp pyramidal features. The average film thickness of sample A is ~10 μm, while that of sample B ranges from ~1 – 5 μm.

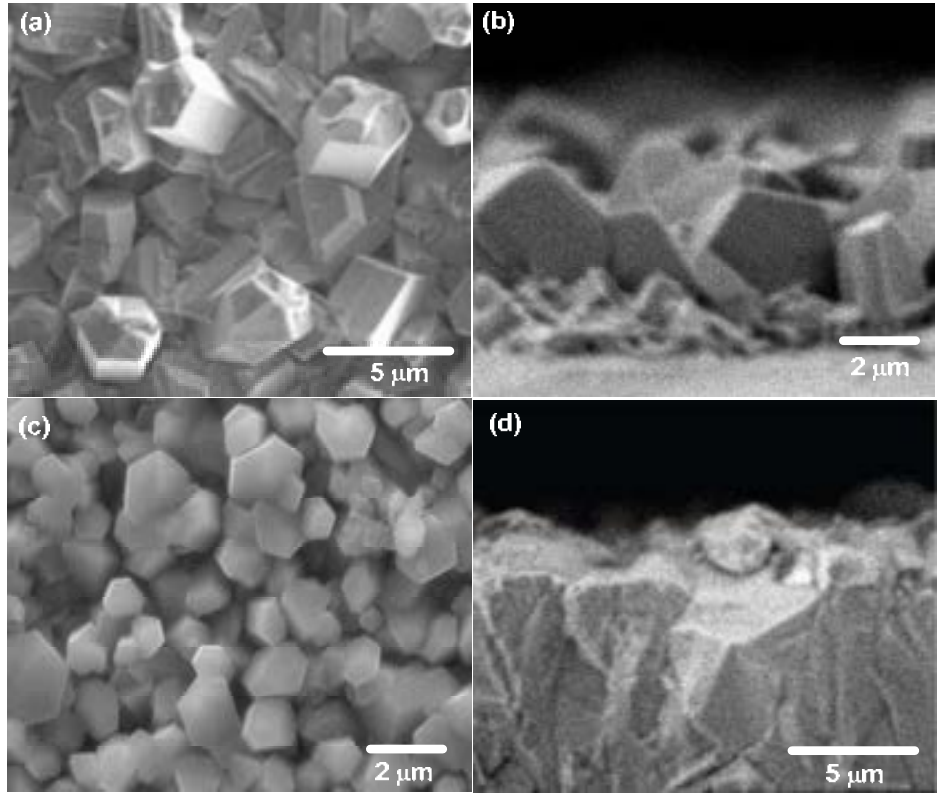

FIG. 6.2. Secondary electron images of GaN film. (a) plan view and (b) cross-section view of sample A, (c) plan view and (d) cross-section view of sample B.

The XRD spectra of both samples are shown in Fig. 6.3. The spectrum of sample A in Fig. 6.3(a) exhibits sharp hexagonal GaN peaks at 32.33°, 34.48° and 36.78°, corresponding to the $(10\bar{1}0)$, $(0002)$, and $(10\bar{1}1)$ planes, respectively. The FWHM of the $(10\bar{1}1)$ peak is 0.230° (824 arcsec). The cubic GaN (002) plane appears at 39.94° with a FWHM of 0.195° (702 arcsec). The presence of this smaller peak indicates cubic inclusions in the primarily hexagonal film of sample A. Due to the overlap of (111) and (0002) diffraction planes between the cubic and hexagonal structures, the unique lattice spacing of the cubic (002) and hexagonal $(10\bar{1}1)$ planes are used here for clear identification of the different phases. The lattice parameters are calculated to be c = 0.520 nm and a = 0.320 nm for the hexagonal GaN and a = 0.451 nm for the cubic GaN, which are in agreement with the reported values for these structures.[78, 79, 108]

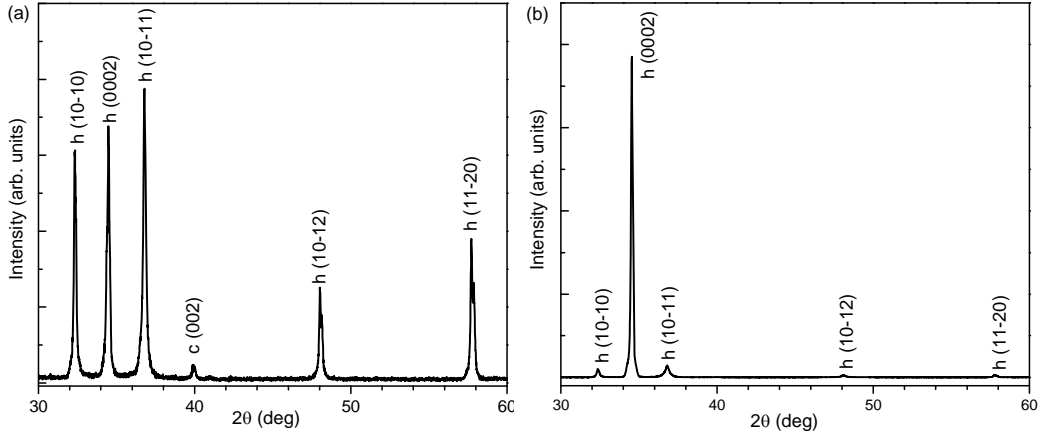

FIG. 6.3.  XRD spectrum of the GaN films. (a) Spectrum of sample A, (b) spectrum of sample B. Miller indices of the planar reflections are labeled corresponding to their phase, either hexagonal (h) or cubic (c).

The spectrum of sample B in Fig. 6.3(b) also exhibits the hexagonal GaN peaks located at 32.38°, 34.55°, and 36.80° corresponding to (10$\bar{1}$0), (0002), and (10$\bar{1}$1) planes, respectively. The FWHM for the (0002) peak is 0.150° (540 arcsec). Lattice parameters are calculated to be a = 0.319 nm and c = 0.519 nm, in agreement with reported values.[78, 79, 108] The intensity of the (0002) peak is ~ 40 times higher compare to the other peak intensities, indicating the preferred c-axis orientation of the GaN film in this sample, as also observed in the SEM images.

Low-temperature (~ 4.2 K) CL spectra of our GaN films are shown in Fig. 6.4. The CL spectrum of sample A, seen in Fig. 6.4(a) shows two luminescence peaks. The high-energy peak consists of two transitions: the first transition is at ~3.466 eV, corresponding to an acceptor-bound exciton (A°X) emission from the free-standing, relaxed, hexagonal GaN layer; while the second transition, located at ~3.491 eV, is attributed to the free exciton (FX) emission of the hexagonal GaN.[112–115] The low energy peak at ~3.278 eV corresponds to the near band edge (NBE) emission of the GaN cubic phase.[109, 113, 115]

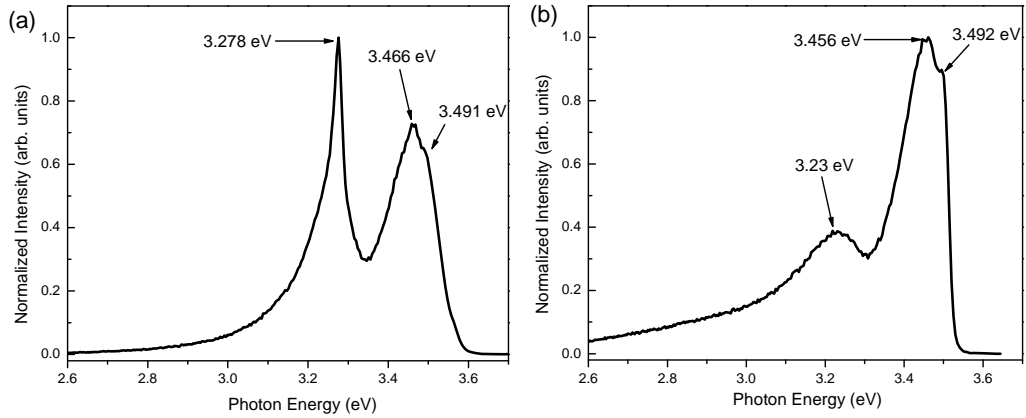

FIG. 6.4.   Low temperature (~4.2 K) cathodoluminescent spectra of a region of the GaN films. (a) CL spectrum of sample A, (b) CL spectrum of sample B.

The CL spectrum of sample B, seen in Fig. 6.4(b) also shows two main luminescence peaks. The high-energy peak consists of the FX emission of the hexagonal GaN at ~3.492 eV and another transitions at ~3.456 eV which is attributed to an acceptor-bound exciton (A°X) emission related to Mg.[112] The low energy peak at ~3.230 eV is attributed to the recombination of an exciton bound to some point defect.[116, 117] Another weak broad emission is present at lower energies, whose origin may also be related to defects.

Spot mode CL analysis, which allows for sampling of the luminescence at a submicron scale, is used on sample A for the identification of the cubic and hexagonal phases. Areas strongly emitting at specific wavelengths are identified by using monochromatic images. By using an excitation of ~3.28 eV (band edge of cubic GaN), we observed highly luminescent regions. When the CL excitation position is set on one of these cubic regions of our GaN film, we obtain the spectrum shown in Fig. 6. 5. This spectrum shows asymmetric characteristics and can be fitted by three Gaussian peaks. The values for the cubic phase of our GaN film are in agreement with those previously reported from low temperature cathodoluminescence measurements of cubic GaN microcrystals[109] and from low temperature photoluminescence of cubic GaN nanowires.[115]

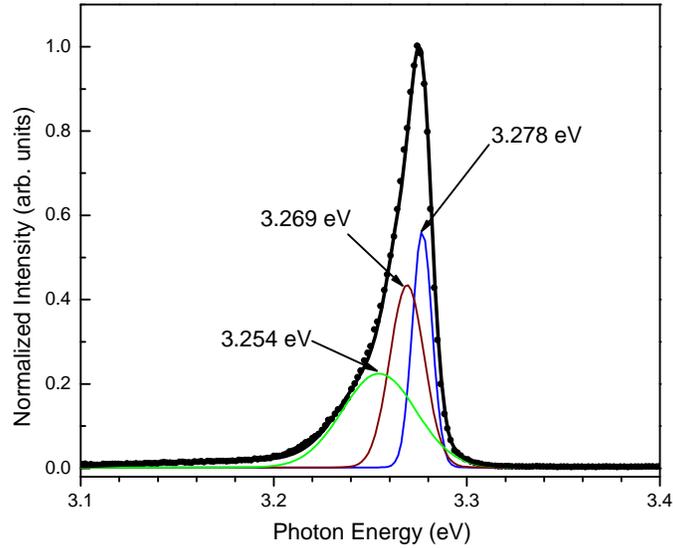

FIG. 6.5.   Low temperature (~4.2 K) spot mode CL spectrum of the cubic GaN region.

The intense narrow peak at ~3.277 eV has a FWHM of 12 meV and has been attributed to free exciton (FX) emission in cubic GaN.[109, 115] The peak at ~3.269 eV is a donor-bound exciton (D°X) emission in cubic GaN.[109, 115] Si or O may be the most probable donors in our unintentionally doped GaN film. The quartz reactor, being in contact with HCl during the growth of our GaN, may act as potential Si and O impurity sources. The third peak in the cubic GaN spectrum is a broad emission centered at ~3.254 eV, which may be defect related or due to acceptor bound exciton (A°X) transitions in cubic GaN,[118] with a Si or C on a N site acting as possible shallow acceptor as already observed in the hexagonal phase of GaN.[119, 120]

A CL spectrum from the hexagonal phase of the GaN film is shown in Fig. 6.6. This spectrum is also asymmetric and consists of four Gaussian peaks located at ~3.491 eV, 3.466 eV, 3.409 eV, and 3.319 eV. The emission peak at ~3.491 eV has a FWHM of 36 meV and is attributed to the FX transition.[121]

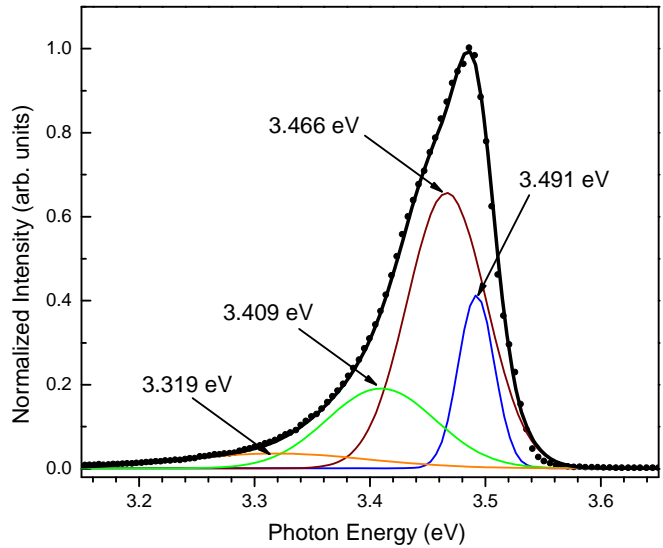

FIG. 6.6.  Low temperature (~4.2 K) spot mode CL spectrum of the hexagonal GaN region.

The emission at ~3.466 eV is related to the exciton bound to a neutral acceptor (A°X). Similar emission has been observed in nominally undoped[122], O-doped[114], and Mg-doped[112, 117] GaN heteroepitaxial layers. Possible sources for the acceptor associated with this transition are the $V_{Ga} - O - H$ complex[114] or a substituting C on a N site[120] due to O or C impurities, respectively. Even though this emission appears in Mg-doped GaN, it may not be Mg-related.[114] The peak at ~3.409 eV is characteristic of excitons bound to basal stacking faults and the very weak peak around 3.319 eV can be linked to excitons bound to prismatic stacking faults.[116, 123, 124] These peaks originate from the stacking faults that form at the cubic-hexagonal interface. The yellow/green luminescence (~ 2.2 – 2.4 eV), commonly observed in hexagonal GaN, is not present in the CL spectra (not shown here).

Fig. 6.7 (a) is a secondary electron image of a region in the film with a corresponding CL spectrum shown in Fig. 6.4(a). The regions associated with peaks of the cubic phase (A) and the hexagonal phase (B) are indicated in the image and correspond to the spectra shown in Fig. 6.5 and 6.6.

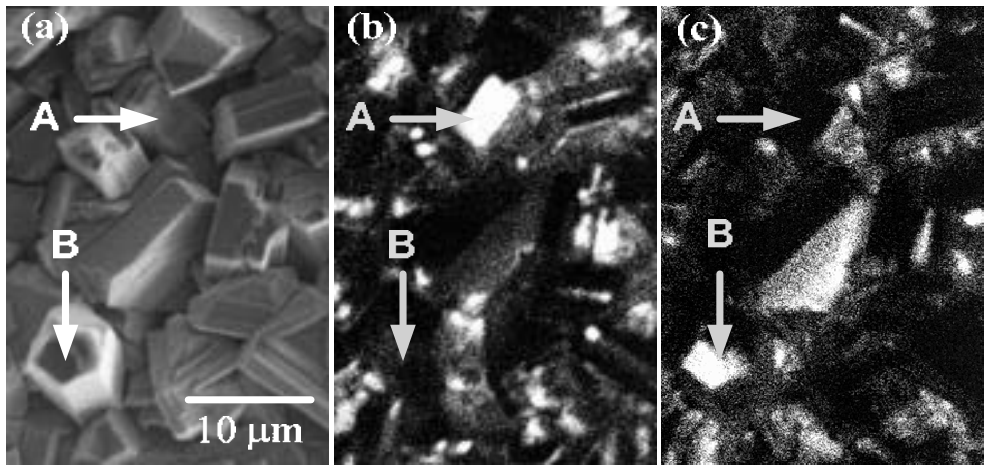

FIG. 6.7.  Morphology and optical properties of GaN film. (a) Secondary electron image from a region of the film with labeled spot mode locations, (b) monochromatic CL image taken at ~378 nm (3.28 eV), and (c) monochromatic CL image taken at ~ 357 nm (3.47 eV).

Fig. 6.7 (b) is a monochromatic image for E = 3.28 eV, where bright regions are attributed to the GaN cubic phase. Fig. 6.7 (c) is a monochromatic image for E = 3.47 eV, corresponding to optical emission from the hexagonal phase of GaN. In these figures the hexagonal and cubic regions appear spatially segregated. Cubic NBE luminescent centers are observed to be ~1-3 μm$^2$ and have well-defined boundaries. From XRD measurements and SE images, the film appears primarily hexagonal. The cubic crystallites contribute to the most intense peak in the luminescence, even though they compose a small portion of the overall film. This may be attributed to the fact that cubic regions, with a narrower band gap, may act as effective radiative recombination centers by drawing electron-hole pairs from the nearby hexagonal film. Different crystalline quality between the two phases may also account for discrepancy in luminescent intensity. Another source of the luminescence seen in CL images at ~3.278 eV may also be from the shallow DAP transitions of the hexagonal phase (~3.27eV), commonly present in undoped GaN and Si/O/Mg doped GaN films.[114, 119, 125] In the GaN film in this study these DAP transitions may be attributed to a shallow donor from substituting Si on a Ga site or possibly substituting O on a N site[114], while a Si or C on a N site may act as an acceptor.[119, 120]

**6.4. DISCUSSION**

It has been reported that epitaxial growth of cubic GaN is favored by the choice of cubic substrates such as GaAs.[108] Hence, it is possible that the FCC structure of the gold substrate droplets may have induced the growth of the cubic phase of the GaN film in this study. However, the fact that the cubic phase was not present on the GaN film grown under the same conditions but at lower temperature indicates that other factors may have influenced its growth.

The fraction of cubic phase inclusions in the hexagonal GaN films has been shown to increase with temperature in HVPE growth.[126] GaN grown at temperatures 700 – 775°C showed mostly hexagonal phase, while growth at 800 – 850°C produced pure cubic layers. The formation of cubic phase is favored under nitrogen deficient conditions.[126, 127]

In this study only the GaN film grown at higher temperature showed cubic inclusions, in agreement with the previously reported results. A model that supports these results is that increasing the temperature causes a decrease in the dissociation rate of ammonia as described in chapter 5 (see Fig. 5.11). This is associated with a decrease in the nitrogen supply during growth. This nitrogen deficiency may create defects such as stacking faults which can result in cubic GaN inclusions. The presence of these stacking faults was confirmed from the PL spectra. The free energy difference between metastable cubic and stable hexagonal phases decreases with the increase of the nitrogen vacancy concentration, making cubic phase more stable. As a conclusion, the formation of the metastable cubic GaN phase is most likely due to non-stoichiometry of the grown film resulting from nitrogen deficient conditions.

## 6.5. CONCLUSIONS

Two films composed of micro-crystals of GaN have been grown by chemical vapor deposition on gold sputtered silica. The films are grown under identical conditions apart from the 100°C difference in substrate temperature, yet there are significant differences on their structural and optical properties. The film grown at lower temperature shows the presence of only hexagonal GaN, while higher temperature growth leads to hexagonal and cubic phases of GaN with the hexagonal phase being predominant as indicated by x-ray diffraction and cathodoluminescence images. Cubic crystallites are found distributed randomly throughout the film, producing sharp exciton luminescence at ~3.278 eV at low temperature. The cubic luminescence peaks have a lower FWHM than their hexagonal counterparts. Even though the cubic phase is a small portion of the overall film, it contributes to the most intense peak in our luminescence studies. The results presented here show that cubic and hexagonal crystallites can coexist, with the cubic phase having a much sharper and stronger luminescence. No long wavelength luminescence has been observed. Controlled growth (temperature, III/V ratio) of the cubic phase GaN crystallites can be of use for high efficiency light detecting and emitting devices.

# CHAPTER 7

# CHARACTERIZATION OF POWDERS GROWN BY AMMONOLYSIS OF A PRECURSOR

## 7.1. INTRODUCTION

InGaN are materials of great importance due to their applications in LEDs and LDs that operate between ultraviolet and visible regions. Several methods are used for the growth of InGaN thin films as mentioned in chapter 3. It is, however, difficult to obtain InGaN with high In content due to high volatility of In at high growth temperatures commonly employed in these methods. The growth of free-standing InGaN in the powder form has been limited up to now with only a few reports in the literature. In this study this possibility is explored by growing III-nitride alloys based on the method described in section 3.4.2. In this chapter the growth and characterization of these powders is reported.

## 7.2. RESULTS AND ANALYSIS

The attempted indium composition $x$ of the samples grown by this method was 0, 0.1, 0.3, 0.5, 0.8, and 1. GaN (x = 0) and InN (x = 1) were grown for comparison. The powders varied in color from light gray (x = 0) to black (x = 1).

The powders have clear hexagonal symmetry as seen in SEM images in Fig. 7.1. Interpenetrated hexagonal particles of GaN powder form round clusters ~200 – 500 nm in diameter as seen in Fig. 7.1 (a). The morphology of the powder for 10% In composition, shown in Fig. 7.1(b), is different from that of pure GaN powder. Individual crystals of hexagonal shape ~1μm in size are present. Smaller size crystals are also seen in the image.

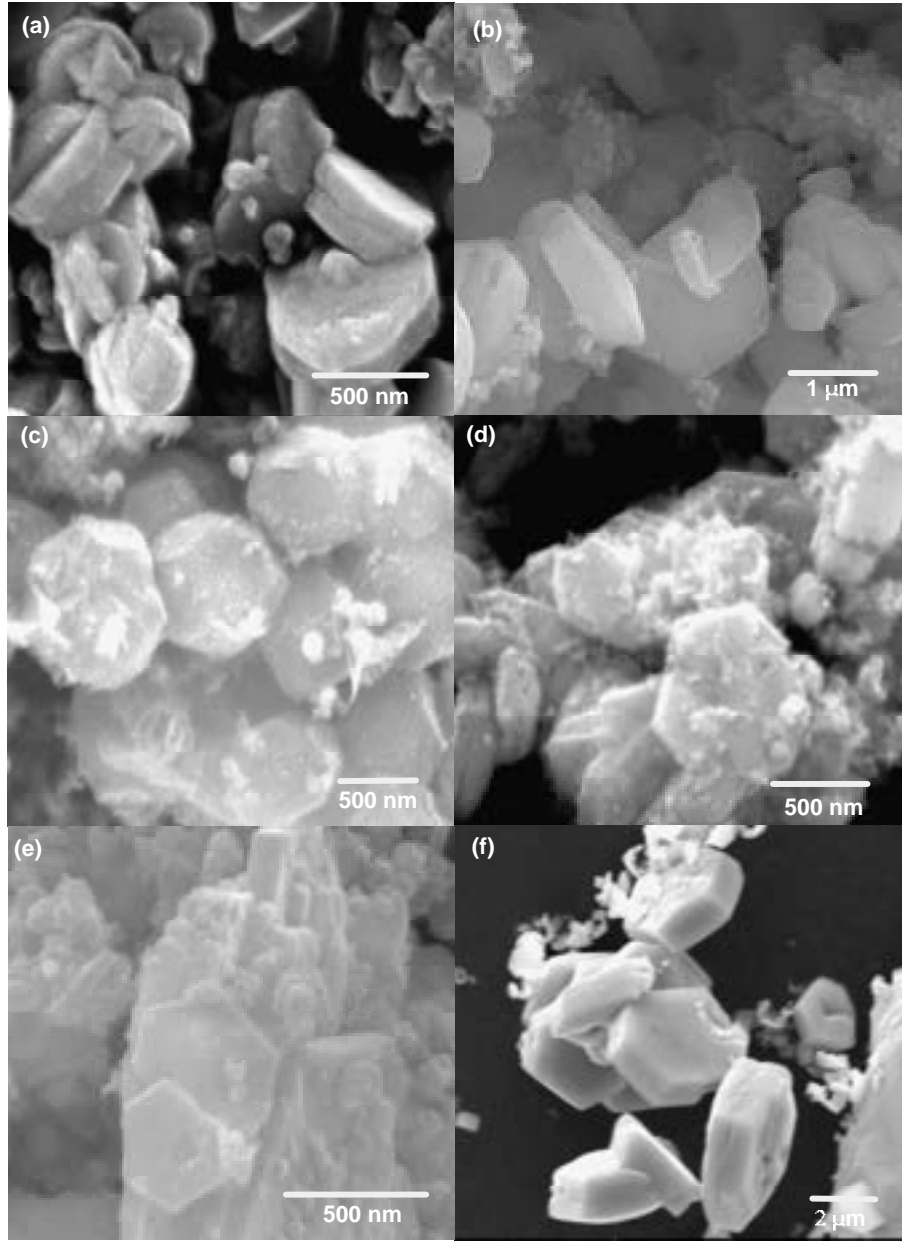

FIG. 7.1.  SEM images of six samples with different In composition. (a) x = 0, (b) x = 0.1, (c) x = 0.3, (d) x = 0.5, (e) x = 0.8, and (f) x = 1.

In Fig. 7.1(c) individual hexagonal submicron particles of the powder with 30 % In composition are observed. The morphology of powders with 50% and 80% In is similar (Fig. 7.1(d) and (e)), in that they both form hexagonal platelets of ~500 nm in diameter. Very fine crystals are also observed. Finally, the SEM image in Fig. 7.1(f) shows that the InN powder forms bigger individual crystals with hexagonal shape that vary in size from ∼ 2 – 4 μm.

The XRD spectra of samples for x = 0 and x = 1 are shown in Fig. 7.2, which indicate the structure of pure hexagonal GaN and InN respectively. The peaks of GaN are broader, characteristic of smaller size crystallites as also observed in SEM images. The peaks for GaN sample are located at ~32.38°, 34.59°, and 36.80° corresponding to $(10\bar{1}0)$, $(0002)$, and $(10\bar{1}1)$ respectively. The diffraction peaks from the same planes of InN sample are located at ~29.19°, 31.36°, and 33.19° respectively. The lattice parameters for GaN are calculated to be a = 0.319 nm and c = 0.519 nm, while for InN the lattice parameters are a = 0.353 nm and c = 0.570 nm, in agreement with reported values for hexagonal structures of these materials.[78, 79, 108]

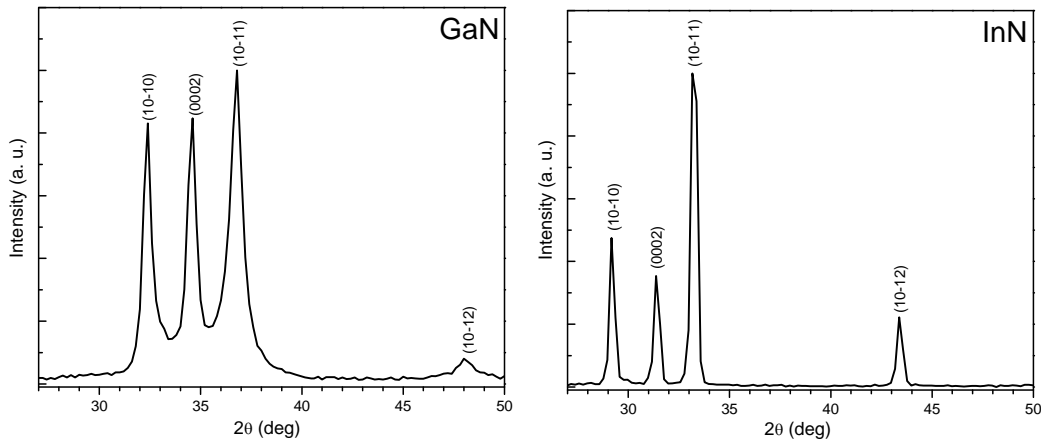

FIG. 7.2.  XRD spectra of GaN and InN grown by ammonolysis of a precursor.

Fig. 7.3 shows the XRD spectra for samples with different indium composition. The spectrum of the sample with 10% In composition seen in Fig.

7.3(a) exhibits peaks of hexagonal GaN and InN. No peaks associated with other phases are observed.

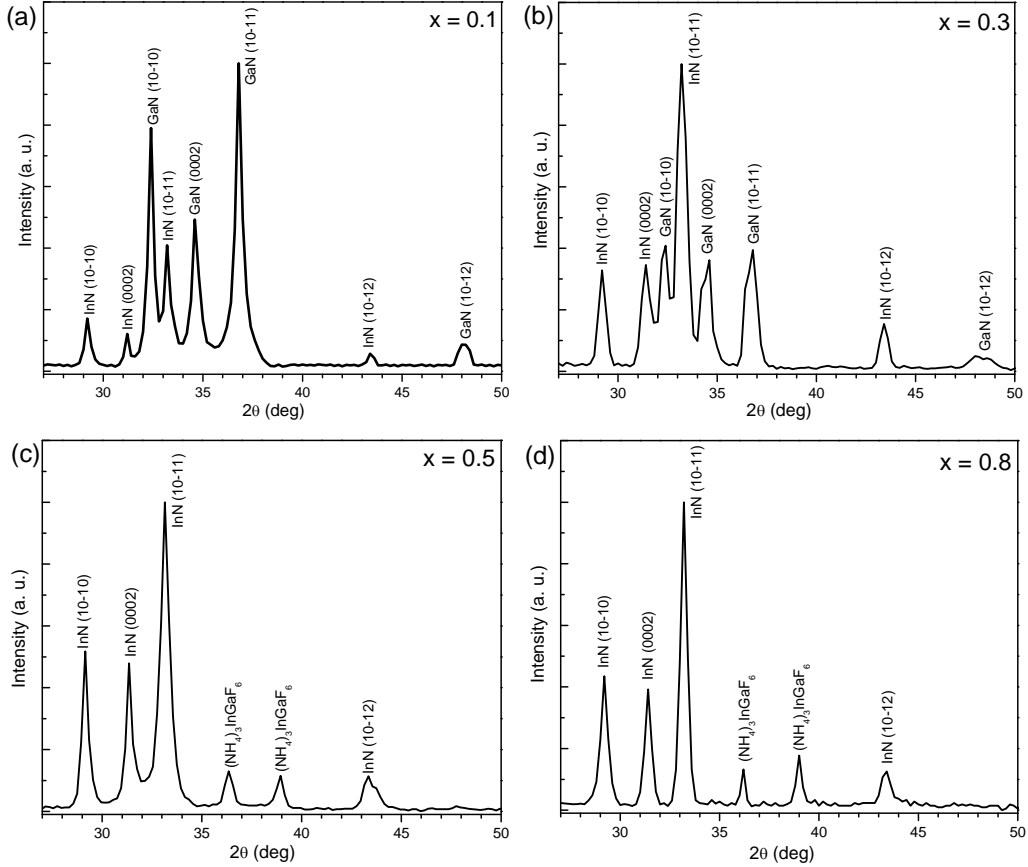

FIG. 7.3. XRD spectra for samples with different indium composition: (a) 10%, (b) 30%, (c) 50%, and (d) 80%.

The lattice parameters of the two structures are calculated to be $a_{GaN} = 0.319$ nm, $c_{GaN} = 0.518$ nm and $a_{InN} = 0.353$ nm, $c_{InN} = 0.570$ nm, which are in agreement with the lattice parameters found for pure GaN and InN. These results indicate a complete phase separation of the InGaN into GaN and InN. From the calculations of the binodal and spinodal curves for the InGaN system in chapter 2, it is predicted that the maximum incorporation of In into the GaN lattice is < 4% at ~630°C. According to the phase diagram the 10 % sample corresponds to a metastable state, so it is possible to grow InGaN with this composition. However, small fluctuations in the growth process can cause the state to be unstable and yield to phase separation.

Fig. 7.3(b) represents the XRD spectrum of the sample with 30% indium composition. The structure of this sample is identified as that of hexagonal InN and GaN. However, the peaks of GaN exhibit asymmetric properties, which is a feature of the presence of InGaN. These peaks are fitted by two Gaussian curves. Lattice parameters are calculated from the position of the peaks given by these curves. They are: $a_{InN}$ = 0.353 nm, $c_{InN}$ = 0.569 nm; $a_{GaN}$ = 0.319 nm, $c_{GaN}$ = 0.518 nm; and $a_{InGaN}$ = 0.322 nm, $c_{InGaN}$ = 0.523 nm. According to Vegard's law, the reduced lattice parameter of InGaN corresponds to ~9% of indium incorporation into the hexagonal GaN. PL measurements at room temperature (see Fig. 7.4) confirm the above results. The PL spectra shows the NBE luminescence of GaN at ~3.37 eV and also the emission of InGaN at ~3.09 eV. Equation (1.1) is employed for the calculation of the band gap of InGaN system. Using 3.4 eV and 0.7 eV for the values of the band gap of GaN and InN respectively, and 1.43 eV as bowing parameter, the composition of InGaN is determined to be ~8%. According to the phase diagram, the 30% indium composition corresponds to an unstable state that leads to phase separation. Separate phases of GaN and InN are observed, as well as InGaN with ~9% indium composition, in agreement with the calculated phase diagram. The dominant peak in the PL spectrum shown in Fig. 7.4 is a broad emission centered at ~1.75 eV, which is the typical red band of GaN unintentionally doped with carbon and oxygen.[117]

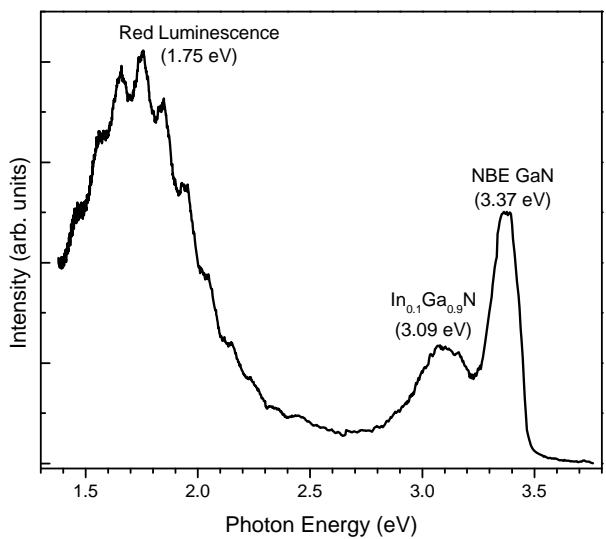

FIG. 7.4.  PL spectrum of the sample with 30% In composition, showing NBE luminescence of GaN and that of InGaN.

The samples with 50% and 80% indium composition exhibit similar peaks in the XRD spectra seen in Fig. 7.3(c) and (d) respectively. Only the InN phase is observed in both spectra, no GaN is present. The lattice parameters of the InN in both samples are calculated to be $a_{InN}$ = 0.353 nm, $c_{InN}$ = 0.570 nm, which are similar to those of pure InN. In addition, two small peaks appear in the spectra between 35° and 40°. These peaks are associated with the tetragonal structure of the precursor $(NH_4)_3InGaF_6$. This leads to believe that not all of the precursor material has been reacted. It also explains the absence of GaN peaks. It is possible that there are some inclusions of GaN in the sample, but their amount might be too small compare to InN in order to appear in the XRD spectra. However, no luminescence was observed from either sample. Both the above compositions correspond to the unstable region of the phase diagram, so phase separation is expected. It is noteworthy that the presence of In metal resulting from InN decomposition is not observed in either sample. This indicates that InN is stable at the growth temperature and method used.

## 7. 3. CONCLUSIONS

The ammonolysis of a precursor was used to grow InGaN powders with different indium composition. High purity hexagonal GaN and InN were obtained. XRD spectra showed complete phase separation for samples with x < 30%, with ~ 9% indium incorporation in the 30% sample. The presence of InGaN in this sample was confirmed by PL measurements, where luminescence from both GaN and InGaN band edge are observed. The growth of samples with higher indium compositions proved to be difficult, with only the presence of InN in the sample. The growth of InGaN by this method was more complicated than expected. Varying parameters like temperature, ammonia flow, and time may be useful to explore. Overall, the difficulty of the growth of InGaN, especially for higher compositions, enforces the validity of the phase diagram produced in chapter 2.

# CHAPTER 8

# CONCLUSIONS AND FUTURE DIRECTION

## 8.1. SUMMARY

High-efficiency blue and green emission is currently achieved using GaN and InGaN thin films, but materials issues still hinder the growth of InGaN alloys, especially for high In composition. As such, the efficiency decreases for samples with longer wavelength emission, toward green and yellow. Difficulties with the growth temperature and indium incorporation call for novel growth approaches. The research presented in this dissertation has attempted to better understand the fundamental thermodynamic and kinetic limitations to producing III-nitride free-standing material.

## 8.2. THERMODYNAMICS OF InGaN PHASE SEPARATION

The thermodynamics of phase separation for the InGaN ternary system has been analyzed using a strictly regular solution model. Graphs of Gibbs free energy of mixing were produced for a range of temperatures. Binodal and spinodal decomposition curves show the stable and unstable regions of the alloy in equilibrium. The interaction parameter used in the analysis is obtained from a strain energy calculation using the valence force field model. According to the calculated results, the critical temperature for InGaN is found to be 1236°C. This suggests that at typical growth temperatures of 800 – 1000°C a wide unstable two-phase region exists in InGaN. Phase separation in epitaxially grown InGaN has been experimentally reported to occur at an indium composition of more than 20%. The predicted temperature and composition region for phase separation are consistent with the experimentally observed difficulties in achieving high In content InGaN required for green emitting devices. These difficulties could be addressed by studying the growth and thermodynamics of these alloys. Knowledge of thermodynamic phase stabilities and of pressure – temperature – composition phase diagrams is important for an understanding of the boundary conditions of a variety of growth techniques. The above study applies only to growth processes under thermal equilibrium conditions and many techniques employed for the growth of InGaN use conditions away from equilibrium. Understanding phase diagrams and the degree of departure from the equilibrium state which corresponds to the so called 'driving force', is useful for these growth processes.

## 8.3. PROPERTIES OF POWDERS GROWN BY Ga-In MELT

The growth of InGaN powders was attempted by using the reaction of molten Ga-In alloy with ammonia in a horizontal tube reactor. X-ray diffraction measurements indicate the hexagonal structure of GaN without any InN or InGaN alloy present. Characterization of the samples by PL shows the NBE emission of GaN, with a weak broader deep level emission and intensity that increases with the indium content in the molten alloy. The growth of InN in our samples, either as a separate phase or as an alloy with GaN, was proven to be unsuccessful. This can be attributed to the low nitrogen pressures used compared to the equilibrium vapor pressure of nitrogen and also to the low decomposition temperature of solid InN. Kinetic and thermodynamic considerations can be used to explain the growth process.

**8.3. COEXISTENCE OF CUBIC AND HEXAGONAL GaN**

Two films composed of micro-crystals of GaN have been grown by chemical vapor deposition on gold sputtered silica at ~800°C and 900°C. XRD and CL measurements show that the film grown at the lower temperature is pure hexagonal GaN, while the one grown at higher temperature contains hexagonal and cubic phases of GaN with the hexagonal phase being the predominant one. Cubic crystallites are found distributed randomly throughout the film, producing sharp exciton luminescence at ~3.278 eV at low temperature. The cubic luminescence peaks have a lower FWHM than their hexagonal counterparts. Even though the cubic phase is a small portion of the overall film, it contributes to the most intense peak in the luminescence studies. The results presented here show that cubic and hexagonal crystallites can coexist, with the cubic phase having a much sharper and stronger luminescence. The formation of the metastable cubic GaN phase is most likely due to non-stoichiometry of the grown film resulting from nitrogen deficient conditions. Controlled growth of the cubic phase GaN crystallites can be of use for high efficiency light detecting and emitting devices.

**8.4. GROWTH OF InGaN POWDERS BY AMMONOLYSIS OF A PRECURSOR**

The ammonolysis of a precursor was used to grow InGaN powders with different indium composition. GaN and InN with pure hexagonal phase were obtained. Complete phase separation was observed for samples with x = 10% and 30%, with ~ 9% indium incorporation in the latter sample. The presence of InGaN in this sample was confirmed by PL measurements, where luminescence from both GaN and InGaN band edge are observed. A red band luminescence was the dominating peaks in the PL spectrum, indicating the presence of impurities like C and/or O in the sample. The growth of samples with higher indium compositions proved to be difficult, with only the presence of InN in the sample. Nonetheless, by controlling parameters like temperature and time may lead to successful growth of this III-nitride alloy by this method.

## 8.5. SUGGESTED FUTURE RESEARCH DIRECTIONS

Achieving high quality InGaN is the main challenge in solid state lighting. The growth of InGaN is hindered by the chemistry, indium incorporation, nitrogen supply, and by the alloy stability. Differences in equilibrium nitrogen vapor pressures also make it difficult to incorporate indium at high growth temperatures (preferred for GaN) and crystalline quality must be sacrificed at low temperatures for indium incorporation. The relationship between growth temperature, alloy composition, and N/III ratio will be crucial for improving crystalline quality. Developing high quality $In_xGa_{1-x}N$ alloys over the entire compositional range is beneficial for a variety of applications such as high efficiency solar cells, terahertz electronics, communications, and optoelectronics. The work shown in this dissertation explores the limits in the fundamental science and technology of solid state lighting. This work demonstrates the highly complicated nature of the growth of InGaN. Nonetheless, the specificity of growth aspects pertaining to this material system offers an added degree of excitement into the research of this nitride semiconductor.

Future work will consist on finding ways for synthesis and stabilization of high [In] alloys and in overcoming the polarization effects. Most likely, successful growth of InGaN with the desired [In] composition will allow the production of quantum wells (QW) with slightly lower [In]. The configuration of an ideal green LED structure is shown in Fig. 8.1.

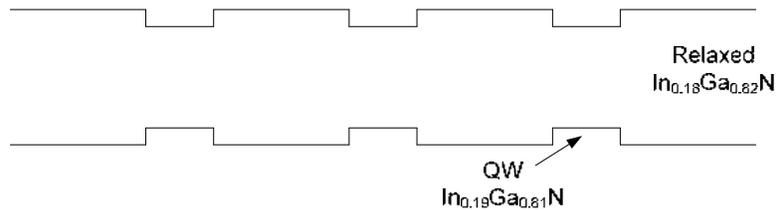

FIG. 8.1. Ideal configuration for high efficiency LED operating in the green region. A relaxed InGaN film (x = 0.18) with an active region of QWs (x = 0.19) minimizing strain and piezoelectric fields.

This type of structure will minimize lattice mismatch and the resulting piezoelectric effects. The challenge is in finding growth methods to produce InGaN and suitable growth orientations that can allow relaxation of lattice mismatch at heterojunctions.